\documentclass[a4paper,preprint,superscriptaddress,preprintnumbers,nofootinbib]{revtex4}

\usepackage[dvipdf,dvips,dvipdfmx]{graphicx}
\usepackage{amsmath}
\usepackage{amssymb}
\usepackage{float}
\usepackage{wrapfig}
\usepackage{bm}
\usepackage{color}
\usepackage{comment}

\allowdisplaybreaks

\newcommand{\be}{\begin{eqnarray}}
\newcommand{\ee}{\end{eqnarray}}

\newcommand{\del}{\partial}

\newcommand{\al}[1]{\begin{align}#1\end{align}}
\newcommand{\nn}{\nonumber\\}

\newcommand{\pmat}[1]{\begin{pmatrix}#1\end{pmatrix}}

\newcommand{\fn}[1]{\!\left(#1\right)}

\newcommand{\bra}{\langle}
\newcommand{\ket}{\rangle}
\newcommand{\Lag}{\mathcal L}

\newcommand{\p}{\partial}

\catcode`\@=11

\def\lsim{\mathrel{\mathpalette\@versim<}}
\def\gsim{\mathrel{\mathpalette\@versim>}}
\def\@versim#1#2{\vcenter{\offinterlineskip
\ialign{$\m@th#1\hfil##\hfil$\crcr#2\crcr\sim\crcr } }}
\catcode`\@=12

\begin{document}

\title{Non-perturbative electroweak-scalegenesis\\
on the test bench of dark matter detection}

\author{Jisuke \surname{Kubo}}
\email{jik@hep.s.kanazawa-u.ac.jp}
\affiliation{Institute for Theoretical Physics, Kanazawa University, Kanazawa 920-1192, Japan}

\author{Qidir Maulana Binu \surname{Soesanto}}
\email{binu@hep.s.kanazawa-u.ac.jp}
\affiliation{Institute for Theoretical Physics, Kanazawa University, Kanazawa 920-1192, Japan}
\affiliation{Department of Physics, Faculty of Science and Mathematics, Diponegoro University, Tembalang, Semarang, Jawa Tengah, 50275, Indonesia}

\author{Masatoshi \surname{Yamada}}
\email{m.yamada@thphys.uni-heidelberg.de}
\affiliation{Institut f\"ur Theoretische Physik, Universit\"at Heidelberg, Philosophenweg 16, 69120 Heidelberg, Germany}

\preprint{KANAZAWA-17-12}

\begin{abstract}
We  revisit  a recently proposed  scale invariant extension of the standard model, in which the scalar bi-linear condensate in a  strongly interacting hidden sector dynamically breaks scale symmetry, thereby triggering  electroweak symmetry breaking.
Relaxing the previously made assumption on $U(N_f)$ flavor symmetry
we find  that the presence of the would-be dark matter candidate 
opens a new annihilation process of dark matter  at finite temperature, such that 
 the model can satisfy  stringent  constraints  of the future experiments of  the dark matter direct detection.

\end{abstract}
\maketitle

\section{Introduction}
What is the origin of mass?
This question has attracted a lot of interests as a big mystery in elementary particle physics.
It has been established by the Large Hadron Collider (LHC) 
\cite{Aad:2012tfa,Chatrchyan:2012xdj}
that there exists a scalar particle, namely, the Higgs boson,
which, as a result of the spontaneous symmetry breaking,
gives the particles in the standard model (SM) a finite mass.
It is, however, unknown how the Higgs field acquires a finite vacuum expectation value.  This is still an open question for a deeper  understanding of the origin of the mass of the SM particles.

The Higgs boson mass parameter is the only dimensionful parameter and  breaks 
the scale invariance in  the SM. 
Its breaking is soft and from this reason
Bardeen \cite{Bardeen:1995kv} argued that
``the SM does not, by itself, have a fine tuning problem
due to the approximate scale invariance of the perturbative expansion''.\footnote{
This fact within the renormalization group was discussed by Wetterich~\cite{Wetterich:1983bi}.
}
The recent idea of a scale invariant extension of the SM in fact
goes back to this observation of Bardeen.
Since the scale invariant classical SM action does not provide
the EW scale, 
it has to be generated by quantum effects.
We here call it ``scalegenesis".

A possible way to realize  scalegenesis in perturbation theory
is the Coleman--Weinberg mechanism~\cite{Coleman:1973jx}, 
where the origin of scale is  the renormalization scale
that has to be introduced unless scale anomaly
is cancelled.
This mechanism cannot, however, yield a stable EW vacuum 
in the SM without contradicting with the observed Higgs boson mass.
Therefore, extensions of the Higgs sector are required.
Along this line of thought,
numerous studies have been dedicated to explain the origin of the EW vacuum.

An alternative approach to realize  scalegenesis 
relies on non-perturbative dynamics.
As is well known, in Quantum chromodynamics (QCD),
 whose action is given as a scale invariant form except 
 for the current quark mass term, a non-trivial vacuum 
 is generated by the strong dynamics of non-Abelian gauge interactions in the infrared energy regime. 
The classically scale invariant extensions of the SM
based on the hidden QCD and their phenomenological implications have been recently  discussed in \cite{Hur:2011sv,Heikinheimo:2013fta,Holthausen:2013ota,Kubo:2014ida,Heikinheimo:2014xza,Carone:2015jra,Ametani:2015jla,Haba:2015qbz,Hatanaka:2016rek,Ishida:2017ehu,Haba:2017wwn,Haba:2017quk,Tsumura:2017knk,Aoki:2017aws}.

In this paper, we consider  scalegenesis realized by 
another non-perturbative dynamics.
We introduce a scale invariant hidden sector,
which is described by 
an $SU(N_c)$ non-Abelian gauge theory
coupled with
$N_f$ complex scalar fields $S_i$ in the fundamental representation of $SU(N_c)$, where the index $i$ denotes the flavor species.
Due to the strong dynamics in the hidden sector, the scalar bi-linear condensate $\langle S_i^\dagger S_j\rangle$ forms,
 triggering the EW symmetry breaking via the Higgs portal coupling $\lambda_{HS,ij}S_i^\dagger S_jH^\dagger H$
 ~\cite{Kubo:2014ova,Kubo:2015cna}.
That is, the dynamical scale symmetry breaking takes place in the hidden sector.
Even though analytic treatments of non-perturbative dynamics are highly complicated, several approaches are available:
One of the possibilities  is an effective theory approach to the non-perturbative dynamics.
Indeed, the Nambu--Jona-Lasinio (NJL) model~\cite{Nambu:1961tp,Nambu:1961fr} has been successfully 
employed  to understand the dynamical chiral symmetry breaking in QCD. It seems obvious that
the basic idea of the NJL model in QCD
can be applied to formulate an effective theory that describes the dynamical scale symmetry breaking.
The first attempt was made in Ref.~\cite{Kubo:2015cna},
in which  the hidden sector is 
effectively described by a scale invariant scalar field theory.
Since the $U(N_f)$ flavor symmetry is unbroken
by the scalar bi-linear condensate
(i.e., $\langle S_i^\dagger S_j\rangle \propto \delta_{ij})$,
the excited states above the vacuum 
with $\langle S_i^\dagger S_j\rangle\neq 0$
are stable and can be identified with
a weakly interacting massive particle (WIMP) dark matter (DM).
The DM relic abundance $\Omega h^2$ and
its spin-independent cross section off the nucleon
$\sigma_{\rm SI}$ 
have been computed   by using the mean-field approximation~\cite{Kubo:2015cna}.
It has been  found there that 
$\sigma_{\rm SI}$ of the model is bounded from below
and is just at the border of the experimental 
upper bound of XENON100~\cite{Aprile:2012nq,Aprile:2013doa}.
Since then there have been  progresses in experiments
\cite{Akerib:2012ys,Aprile:2015uzo,Cao:2014jsa}, so that the minimal model
may be running  into problems with experimental constraints
in  future.
The reason why $\sigma_{\rm SI}$ 
cannot be made small while 
maintaining a correct value of $\Omega h^2$
is that the portal coupling $\lambda_{HS}$  acts on  $\Omega h^2$ and
$\sigma_{\rm SI}$ in an opposite direction.
Therefore, as long as the $U(N_f)$ flavor symmetry is intact,
we cannot avoid this problem.

The main feature of the modified model presented in the present work
is that the $U(N_f)$ flavor symmetry
is explicitly broken by the quartic scalar couplings.
Specifically, we consider the case of the $U(2)$ flavor symmetry, which is broken by
 the quartic scalar couplings down to $U(1)\times U(1)$.
 In the  $U(1)\times U(1)$ invariant model 
 there is one complex scalar for the DM candidate, while there are  three real stable scalars in the  $U(2)$ invariant model.
 The benefit of the explicit breaking 
 of $U(2)$ is that due to the presence of
 the would-be DM candidate 
 (the third scalar in the $U(2)$ invariant model)
 a new annihilation process for DM at finite
 temperature becomes available,
 which is independent of $\lambda_{HS}$.

 In the following section we  start by modifying the minimal model and elucidate our mean field approximation to the strong dynamics,
 which is successively applied to compute effective interactions
of  DM in section IV. They are finally used to obtain 
 $\Omega h^2$ and $\sigma_{\rm SI}$ in section V.
The last section is devoted to  our conclusion.
 
\section{The model}\label{models}
\noindent 
We extend the classical scale invariant extension of the 
SM, which has been studied in \cite{Kubo:2015cna,Kubo:2015joa,Kubo:2016kpb}.
The hidden sector, in which 
the  EW scale originates,
 is described by
an $SU(N_c)$ gauge theory with
 the scalar fields $S_i^{a}$ ($a=1,\dots,N_c$, $i=1,\dots,N_f$) in
the fundamental representation of $SU(N_c)$.
Instead of the $U(N_f)$ flavor symmetry, which was assumed in 
\cite{Kubo:2015cna,Kubo:2015joa,Kubo:2016kpb},
we assume here  only   $U(1)^{N_f}$ symmetry.
The total $U(1)^{N_f}$ invariant Lagrangian for the extended model is given by
\al{
{\cal L}_{\rm H} &=
-\frac{1}{2}\mbox{tr}\,\{F_{\mu\nu}F^{\mu\nu}\}
+([D^\mu S_i]^\dag D_\mu S_i)
-\hat{\lambda}_{S_{ij}}(S_i^\dag S_i) (S_j^\dag S_j)\nn
&\qquad -\hat{\lambda'}_{S_{ij}}
(S_i^\dag S_j)(S_j^\dag S_i)
+\hat{\lambda}_{HS_{i}}(S_i^\dag S_i)H^\dag H
-\lambda_H ( H^\dag H)^2+{\cal L}'_{\mathrm{SM}},
\label{LH}
}
where the parenthesis $(~~)$ stands for $SU(N_c)$ invariant products, $D_\mu S_i = \partial_\mu S_i -ig_H G_\mu S_i$, $G_\mu$ is the matrix-valued $SU(N_c)$ gauge field, $F_{\mu\nu}$ is the field strength tensor of $G_\mu$, the SM Higgs doublet field is denoted by $H$, and
${\cal L}'_{\mathrm{SM}}$ contains the SM gauge and Yukawa interactions.
The scale-invariance violating Higgs mass term is absent in \eqref{LH}.

Our basic assumption is as before that the origin of 
the  EW scale  is  a scalar-bilinear condensation, 
\al{
\langle (S^\dag_i S_j)\rangle =
\langle \sum_{a=1}^{N_c} S^{a\dag}_i S^a_j\rangle=f_{ij},
\label{condensate}
}
which forms due to the 
$SU(N_c)$ gauge interaction and 
triggers the EW symmetry breaking through the Higgs portal coupling
$\hat{\lambda}_{HS_i}$.
The condensation \eqref{condensate} will also generate 
 the mass term (constituent mass) for $S_i$ dynamically.
In \cite{Kubo:2015cna,Kubo:2015joa,Kubo:2016kpb} we have proposed to describe this non-perturbative phenomena of condensation approximately by using an effective theory.
As in the case of the NJL theory, which can effectively describe the dynamical
chiral symmetry breaking in QCD,  the effective Lagrangian  contains only
the ``constituent'' scalar fields $S_i^{a}$.
Furthermore,  in writing down  the effective Lagrangian at the tree level,
we have ignored the presence of scale anomaly,
because its breaking is only logarithmic and cannot generate a mass term.
That is, the breaking of scale invariance is hard, but not soft.
Here we restrict ourself to the minimal model, i.e., to $N_f=2$.
The effective Lagrangian  then can be written as
\al{
{\cal L}_{\rm eff} &=
 ([\partial^\mu S_i]^\dag \partial_\mu S_i)-
\lambda_{1}(S_1^\dag S_1) (S_1^\dag S_1)-
\lambda_{2}(S_2^\dag S_2) (S_2^\dag S_2)
-\lambda_{12}
(S_1^\dag S_1)(S_2^\dag S_2)\nn
&\quad -\lambda'_{12}(S_1^\dag S_2) (S_2^\dag S_1)
+\lambda_{HS_i}(S_i^\dag S_i)H^\dag H
-\lambda_H ( H^\dag H)^2,
\label{Leff}
}
where all coupling constants
are positive, and
\al{
\lambda_1&=\lambda_{S_{11}}+\lambda'_{S_{11}},&
\lambda_2&=\lambda_{S_{22}}+\lambda'_{S_{22}},&
\lambda_{12}&=\lambda_{S_{12}}+\lambda_{S_{21}},&
\lambda'_{12}&=\lambda'_{S_{12}}+\lambda'_{S_{21}}.&
}
The effective Lagrangian ${\cal L}_{\rm eff}$  
is the most general form which is
consistent with 
the global $SU(N_c)\times U(1)^{N_f}$ symmetry and the classical scale invariance.\footnote{
We have suppressed ${\cal L}'_{\mathrm{SM}}$ as well as 
the kinetic term for $H$ in \eqref{Leff},
because they  do not play any important role for our 
discussions here.}
Needless to say that  ${\cal L}_{\rm eff}$ 
has the same global symmetry as  ${\cal L}_{\rm H}$ 
even at the quantum level.
Note also that, though the structure of the quartic couplings
of $S$ in  $ {\cal L}_{\rm eff}$ is the same as that in 
$ {\cal L}_{\rm H}$, the couplings
$\hat{\lambda}_{S_{ij}}, \hat{\lambda'}_{S_{ij}}$, and $\hat{\lambda}_{HS_i}$ 
in ${\cal L}_{\rm H}$ are not the same as
$\lambda_{S}, \lambda'_{S}$, and $\lambda_{HS}$ in $ {\cal L}_{\rm eff}$,
because the unhatted ones are dressed by
the $SU(N_c)$ gauge field contributions.

\section{Physical quantities within mean field approximation}\label{Mean field }
We employ the auxiliary field method to investigate the vacuum structure of the effective Lagrangian ${\cal L}_{\rm eff}$.
In particular, we here would like to see that the non-perturbative dynamics described by ${\cal L}_{\rm eff}$ actually realizes the non-trivial vacuum \eqref{condensate} in the hidden sector.
To this end, we introduce auxiliary fields $f_i$ and $\phi^{\pm}$
($\phi^+=(\phi^-)^*$)
and add 
\al{
{\cal L}_{\rm ax}=
\lambda_1 f_{1}^2+\lambda_2 f_{2}^2+
\lambda_{12} f_{1} f_{2}+
\frac{1}{2}\lambda'_{12} \phi^+ \phi^-
\label{ax}
}
to the effective Lagrangian \eqref{Leff}.
Note that since the path integrals of $f_i$, $\phi^\pm$ are Gaussian ones, at the tree-level, the contributions from these fields have no effects on the effective theory.
We then shift them according to
\al{
f_{1} &\to f_{1} -(S_1^\dag S_1),&
f_{2}&\to f_{2} -(S_2^\dag S_2),&
\phi^+ &\to \phi^+ -\sqrt{2}(S_2^\dag S_1),&
\phi^- &\to \phi^- -\sqrt{2}(S_1^\dag S_2)&
}
to obtain the mean-field Lagrangian 
\al{
{\cal L}_{\rm MFA}
& =
 ([\partial^\mu S_i]^\dag \partial_\mu S_i) 
 -M_{i0}^2(S_i^\dag S_i)
+ \lambda_1 f_{1}^2 + \lambda_2 f_{2}^2 +
 \lambda_{12} f_{1}f_{2}\nn
 &\quad -\lambda_H(H^\dag H)^2+\frac{\lambda_{12}'}{2}\phi^+\phi^-
-\frac{\lambda'_{12}}{\sqrt{2}}\phi^+(S^\dag_1 S_2)
-\frac{\lambda'_{12}}{\sqrt{2}}\phi^-(S^\dag_2 S_1),
\label{LMFA}
}
where
 \al{
M_{10}^2 &=
2\lambda_1 f_{1}+\lambda_{12} f_{2}-
\lambda_{HS_1} H^\dag H,&
M_{20}^2 &=2\lambda_2 f_{2}+\lambda_{12} f_{1} -
\lambda_{HS_2} H^\dag H.&
\label{cons scalar mass}
}
Note that the mean-field Lagrangian reduces to ${\cal L}_{\rm eff}$
when the tree-level equations of motion for  the auxiliary fields,
$f_{i} =(S_i^\dag S_i),~\phi^+ =\sqrt{2}(S_2^\dag S_1)$,
are plugged into \eqref{LMFA}. 

To proceed with the mean-field approximation, we have to derive
the effective potential $V_\text{MFA}$ for our problem.
By assumption the non-perturbative effect of the original
gauge theory breaks neither the hidden $SU(N_c)$ color symmetry nor
 the $U(1)\times U(1)$ flavor symmetry ,
which means 
that $\bra S_i \ket =0$ and 
$\bra (S_2^\dag S_1) \ket=\bra \phi^+ \ket/\sqrt{2} =
\bra (S_1^\dag S_2) \ket=\bra \phi^- \ket /\sqrt{2} =0$.
Therefore, we ignore the last three terms involving $\phi^\pm$ in \eqref{LMFA} and calculate the 
$V_\text{MFA}$ by integrating out the scalar fields $S$ whose integration is Gaussian.
Then, we find the effective potential:
\al{
V_\text{MFA}\fn{f,H}
&=  - \lambda_1 f_{1}^2 - \lambda_2 f_{2}^2
 -\lambda_{12} f_{1}f_{2}+\lambda_H (H^\dag H)^2 \nn
&\quad
+\frac{N_c}{32\pi^2}M_{10}^4 \ln\frac{M_{10}^2}{\Lambda_H^2}+
\frac{N_c}{32\pi^2}M_{20}^4 \ln\frac{M_{20}^2}{\Lambda_H^2},
\label{effective potential MFAa}
}
where $M_{i0}^2$ are given in \eqref{cons scalar mass}, the ultraviolet divergence was subtracted with the $\overline{\rm MS}$ scheme, and $\Lambda_H=\mu e^{-3/4}$ is a renormalization scale at which the quantum corrections vanish.

We here stress that the scale is generated by quantum effects within the scaleless effective theory \eqref{Leff}.\footnote{
Although in the original gauge theory \eqref{LH}, the non-trivial scale may be generated by its strong dynamics, it is complicated.
Instead, we have attempted to demonstrate that the scale generation by the strong dynamics is realized by the dimensional transmutation {\it \`a la} the Coleman--Weinberg mechanism.}
This scale characterizes the origin of the scales of both the hidden sector and the EW.

The  minima of the effective potential \eqref{effective potential MFAa} 
can be obtained from  the solution of the gap equations\footnote{A similar potential problem has been studied in \cite{Coleman:1974jh,Kobayashi:1975ev,Abbott:1975bn,Bardeen:1983st}.
But they did not study the classical scale invariant case in detail,
and moreover no coupling to the SM was introduced.}
\al{
0&= \frac{\del}{\del f_{i}}V_{\rm MFA}
=\frac{\del}{\del H_l}V_{\rm MFA},&(i,l&=1,2).&
\label{station}
}
The first equation in \eqref{station} yields 
\al{
N_c \bra M_{i0}^2\ket
 \left[ \ln (\bra M_{i0}^2 \ket/\Lambda_H^2)+\frac{1}{2}  \right] 
= 16  \pi^2 \bra f_{i} \ket,\label{VEVf}
}
which implies that $\bra M_{i0}^2\ket=0$ if $\bra f_{i} \ket=0$.
In the case that $\ln (\bra M_{i0}^2 \ket/\Lambda_H^2)+1/2 
<0$, \eqref{VEVf} is inconsistent unless 
 $\bra M_{i0}^2\ket=\bra f_{i} \ket=0$, 
 because $\bra M_{i0}^2\ket$
are $\bra f_{i} \ket$ are not allowed  to be negative.
Then the second equation of \eqref{station} gives
\al{
2\lambda_H \bra H^\dag H\ket =
\lambda_{HS_1}\bra f_{1}\ket+\lambda_{HS_2} \bra f_{2}\ket.
\label{VEVH}
}
Using \eqref{VEVf} and \eqref{VEVH}, we find the 
potential at the minimum:
\al{
 \langle V_{\rm MFA}\rangle
 =-\frac{N_c}{64\pi^2}\left( \bra M_{10}^2\ket^2+
 \bra M_{20}^2\ket^2\right).
 \label{minimum}
 }
From \eqref{VEVH} we see that, if $\bra H^\dag H\ket$
vanishes, $\bra f_{1}\ket$ and $\bra f_{2}\ket$ also
have to vanish, because we assume 
that $\lambda_H, \lambda_{HS_i}$ are positive.
 $\bra H^\dag H\ket=0$ cannot be at a local minimum unless 
both $\bra f_{i} \ket$ vanish, 
which can be seen from the Higgs mass squared
\al{
m_{h0}^2
&=
6\lambda_H \bra H^\dag H\ket+
\frac{ N_c(\lambda_{HS_1}^2+\lambda_{HS_2}^2) \bra H^\dag H\ket}{8\pi^2}\nn
& \quad -\lambda_{HS_1} \bra f_{1}\ket
-\lambda_{HS_2} \bra f_{2}\ket
+2\bra H^\dag H\ket\left(\lambda_{HS_1}^2
 \frac{\bra f_{1}\ket}{ M_1^2}+
\lambda_{HS_2}^2
 \frac{\bra f_{2}\ket}{ M_2^2}\right)\label{mh0}\\
& \to -(\lambda_{HS_1}\bra f_{1}\ket
+\lambda_{HS_2}\bra f_{2}\ket)
< 0 ~\mbox{as}~\bra H^\dag H\ket \to 0,\nonumber
}
where we have not used \eqref{VEVH}.
Therefore, $\bra H^\dag H\ket =0$ is possible only if 
$\bra f_{1}\ket =\bra f_{2}\ket=0$, which is consistent with \eqref{VEVH}.
Furthermore, one can convince oneself that eqs.\,\eqref{VEVf} and \eqref{VEVH} cannot be simultaneously satisfied
if one of $\bra f_{i} \ket$ vanishes, unless we make a precise fine-tuning of the quartic coupling constants.
From the discussions above we may therefore conclude
that, as long as  $\ln (\bra M_{i0}^2 \ket/\Lambda_H^2)+1/2 
> 0$ is satisfied, the non-vanishing VEV of $H$ and $f_{i}$ correspond to the true minimum of the
potential \eqref{effective potential MFAa}.\footnote{At finite temperature, the scale invariance is explicitly 
broken, and a Higgs mass term is effectively generated.
As a consequence,
$\bra f_{i}\ket \neq 0$ but
 $\bra H^\dag H\ket =0$ can become possible \cite{Kubo:2015joa}.}

To proceed with our mean-field approximation, we introduce
fluctuations about the mean-field vacuum
\eqref{VEVf}--\eqref{minimum} as
\al{
f_{i} &= \bra f_{i}\ket +\sigma_i.
\label{sigma}
}
Note that $\phi^\pm$ are also fluctuations
and also that the canonical dimension of $\sigma_i$ and 
$\phi^\pm$ is two.
Similarly, we expand  the Higgs doublet around the 
vacuum value as
\al{
H&= \frac{1}{\sqrt{2}}
\pmat{
\chi^1+i\chi^2\\
v_h+h +i\chi^0},&
\frac{v_h}{\sqrt{2}}&=(\bra H^\dagger H\ket)^{1/2},&
\label{field expansion}
}
where $\chi^i$ are would-be Nambu-Goldstone fields and we will suppress them in the following discussions.
Then the mean-field Lagrangian \eqref{LMFA} can be
finally  written as
\al{
\Lag_\text{MFA}&=
 ([\partial^\mu S_i]^\dag \partial_\mu S_i) 
 -M_i^2(S_i^\dag S_i)+\frac{\lambda_{12}'}{2} \phi^+\phi^-
  + \lambda_1 \sigma_1^2 + \lambda_2 \sigma_2^2 +
 \lambda_{12} \sigma_1\sigma_2\nn
&\quad 
+ \lambda_1 f_1^2 + \lambda_2 f_2^2 +
 \lambda_{12} f_1f_2 -(2\lambda_1 \sigma_1+
 \lambda_{12} \sigma_2)(S_1^\dag S_1)\nn
 & \quad -
 (2\lambda_2  \sigma_2+
 \lambda_{12} \sigma_1)(S_2^\dag S_2)
 -\frac{\lambda_{12}'}{\sqrt{2}}\phi^+(S^\dagger_1 S_2)
-\frac{\lambda_{12}'}{\sqrt{2}}\phi^-(S^\dagger_2 S_1)
 \nn
&\quad +\frac{\lambda_{HS_i}}{2}(S_i^\dag S_i) h (2v_h +h)
-\frac{\lambda_H}{4}h^2(6v_h^2 +4v_h h + h^2),
\label{MFA2ap}
}
where 
\al{
M_1^2 &= \bra M_{10}^2\ket =2\lambda_1 \bra f_{1}\ket +\lambda_{12} \bra f_{2}\ket-\frac{
\lambda_{HS_1}}{2} v_h^2,\\
M_2^2 &=\bra M_{20}^2\ket = 2\lambda_2 \bra f_{2}\ket+\lambda_{12}\bra f_{1}\ket -\frac{\lambda_{HS_2}}{2} v_h^2.
\label{cons scalar mass 2}
}
At this level  
the mean fields $\sigma_i$ and $\phi^{\pm}$ are classical fields, but we reinterpret them as quantum fields after their kinetic terms are generated at the loop level.
More specifically, the auxiliary fields, $\sigma_i$ and $\phi^\pm$, are not dynamical in the Lagrangian at the classical level~\eqref{MFA2ap}.
As will be seen in the next subsection, these fields become dynamical by integrating out the fundamental fields $S_i$.
Note that within the present effective model approach to dynamical scale symmetry breaking described by \eqref{LH}, the confinement effects cannot be taken into account.

Here, we briefly introduce the one-loop contribution from the SM sector 
to the effective potential~\eqref{effective potential MFAa} and evaluate the correction to the Higgs mass~\eqref{mh0}.
The  one-loop contribution to the effective potential is calculated as
\al{
V_\text{CW}\fn{h} = \sum_{I=W,Z,h} \frac{n_I}{2}\int \frac{d^4k}{(2\pi)^4} \ln\fn{k^2 + m_I^2\fn{h}}
- \frac{n_t}{2}\int \frac{d^4k}{(2\pi)^4} \ln\fn{k^2 + m_t^2\fn{h}}+\mbox{c.t.},
\label{one loop potential}
}
where $n_I$ ($I=W,Z,t,h$) is the degrees of freedom of the corresponding  particle, i.e., $n_W=6$, $n_Z=3$, $n_t=12$ and $n_h=1$,
and c.t. denotes the counter terms. 
We work in the Landau gauge, and the contributions from the would-be Goldstone bosons in  the Higgs field have been neglected.
We employ the dimensional regularization in order to respect scale invariance and choose the counter terms such that the following normalization conditions with $v_h=246$ GeV are satisfied:
\al{
V_{\rm CW}\fn{h=v_h} &=0,&
\frac{d V_{\rm CW}\fn{h} }{d h}\bigg|_{h=v_h}&=0.&
\label{normalization}
} 
Then, we obtain the one-loop corrections \eqref{one loop potential} as the Coleman--Weinberg potential~\cite{Coleman:1973jx} 
\al{
V_{\rm CW}\fn{h} &=
C_0 (h^4-v_h^4)+\frac{1}{64 \pi^2}
\bigg[ ~6 \tilde{m}_W^4 \ln (\tilde{m}_W^2/m_W^2)+
3 \tilde{m}_Z^4 \ln (\tilde{m}_Z^2/m_Z^2)\nn
&\quad
+ \tilde{m}_h^4 \ln (\tilde{m}_h^2/m_h^2)
  -12\tilde{m}_t^4 \ln (\tilde{m}_t^2/m_t^2)~\bigg],
  \label{VCW}
}
where
\al{
C_0 &\simeq -\frac{1}{64 \pi^2 v_h^4}\left(3m_W^4+(3/2)m_Z^4+(3/4) m_h^4-6 m_t^4\right),\label{C0}\\
\tilde{m}_W^2 &=(m_W/v_h)^2h^2,~~~
\tilde{m}_Z^2=(m_Z/v_h)^2 h^2,~~~
\tilde{m}_t^2 =(m_t/v_h)^2 h^2,~~~
\tilde{m}_h^2 =\frac{\p^2 V_\text{MFA}}{\p h^2},
\label{mh2}
}
and $m_I$ (masses given at the vacuum $v_h=246$ GeV) are 
\al{
m_W&=80.4~\mbox{GeV},& m_Z&=91.2~\mbox{GeV}, &m_t&=173.2~\mbox{GeV},& m_h&=125~\mbox{GeV}.&
\label{mass at vacuum v}
}
We find that the Coleman--Weinberg potential  \eqref{VCW} yields a one-loop correction to the Higgs mass squared \eqref{mh0} 
\al{
\delta m_h^2 = \frac{d^2V_\text{CW}}{dh^2}\bigg|_{h=v_h}
 \simeq -16 C_0 v_h^2.
\label{correction to higgs}
}
This correction modifies the Higgs mass \eqref{mh0} slightly.
\subsection{Inverse propagators and masses}
The inverse propagators 
should be computed to obtain 
the masses and the corresponding wave 
function renormalization constants.
We also have to define canonically normalized fields with 
a canonical dimension of one.
To this end, we  integrate out the constituent
scalars $S^a$ and up to and including one-loop order to obtain
the inverse propagators:
\al{
\Gamma_\phi(p^2)
&=\frac{1}{2}\lambda'_{12}
\left[1+
\lambda'_{12} N_c\Gamma(M_1^2,M_2^2,p^2)\right],
\label{gamma}\\
\Gamma_{11}(p^2)
&=2\lambda_{1}\left[1+
2 N_c \lambda_{1}\Gamma(M_1^2,M_1^2,p^2)\right]+
 N_c \lambda_{12}^2\Gamma(M_2^2,M_2^2,p^2),\\
 \Gamma_{22}(p^2)
&=2\lambda_{2}\left[1+
2 N_c \lambda_{2}\Gamma(M_2^2,M_2^2,p^2)\right]+
 N_c \lambda_{12}^2\Gamma(M_1^2,M_1^2,p^2),\\
  \Gamma_{12}(p^2)
&=\lambda_{12}\left[1+
2 N_c \lambda_{1}\Gamma(M_1^2,M_1^2,p^2)+
2 N_c \lambda_{2}\Gamma(M_2^2,M_2^2,p^2)
\right],\\
\Gamma_{h1}(p^2)
&=-v_h\left[
2\lambda_{HS_1}\lambda_{1} N_c \Gamma(M_1^2,M_1^2,p^2)
+\lambda_{HS_2}\lambda_{12} N_c \Gamma(M_2^2,M_2^2,p^2)
\right],\\
\Gamma_{h2}(p^2)
&=-v_h\left[
2\lambda_{HS_2}\lambda_{2} N_c \Gamma(M_2^2,M_2^2,p^2)
+\lambda_{HS_1}\lambda_{12} N_c \Gamma(M_1^2,M_1^2,p^2)
\right],\\
\Gamma_h(p^2)
&=p^2-m_{h}^2+(v_h \lambda_{HS_i})^2
N_c ~\left[\Gamma(M_i^2,M_i^2,p^2)-
\Gamma(M_i^2,M_i^2,0)\right],
}
with $m_{h}^2=m_{h0}^2+\delta m_h^2$, where
$m_{h0}^2$ is given in \eqref{mh0},
$\delta m_h^2$ is the SM correction given in \eqref{correction to higgs}, and 
we defined the loop function,
\al{
\Gamma(M_1^2,M_2^2,p^2)
=\frac{-1}{16\pi^2}
\left(\int_0^1 dx \ln\{ 1-x(1-r)-x(1-x)t\}
+\ln\left[\frac{M_2^2}{\Lambda_H^2 \exp(-3/2)}\right]\right),
\label{loop function}
}
with $r=M_1^2/M_2^2$ and $t=p^2/M_2^2$.
Note that we have included the canonical kinetic term for $H$,
but  its wave function renormalization constant is ignored, 
because it is approximately equal to one within the approximation
here.
Note that the fundamental fields $S_i$ have been integrated out, so that they are no longer fields as degrees of freedom in low energy regimes (below the confinement scale).
Instead, the auxiliary fields associated with the composite fields could behave as dynamical fields with degrees of freedom in low energy regimes.
The  DM mass is  the momentum squared at which the inverse propagator of $\Gamma_\phi\fn{p^2}$ vanishes, i.e.,
\al{
\Gamma_{\phi}(p^2 = {m_{\rm DM}}^2)=0,
\label{zero}
}
and $Z_{\phi}$ (which has a canonical  dimension  of two) can be obtained from
\al{
Z_{\phi}^{-1} &= \frac{d \Gamma_\phi}{d p^2}\bigg|_{p^2=m_{\rm DM}^2}.
}
The Higgs and $\sigma_i$ masses can be similarly  obtained 
from the eigenvalues of the following $h-\sigma$ mixing matrix
\al{
{\bf \Gamma}(p^2) &=
\pmat{
\Gamma_{11}(p^2) &\Gamma_{12}(p^2) & \Gamma_{h1}(p^2)\\
\Gamma_{12}(p^2) & \Gamma_{22}(p^2)  & \Gamma_{h2}(p^2) \\
\Gamma_{h1}(p^2) & \Gamma_{h2}(p^2)  & \Gamma_{h}(p^2) 
}.
\label{mixing}
}
The squared  masses  $m^2_a~(a=H,L,h)$ are  given as the momenta at which $\det {\bf\Gamma}(p^2)$ becomes zero, where we assume that
\al{
m_H > m_L > m_h.
}
Further, the wave function renormalization constants can be computed in the following way. We first compute the squared masses from
$\det {\bf\Gamma}(p^2) =0$.
Then we diagonalize ${\bf\Gamma}(p^2)$ at each $p^2=m_a^2$
and  denote the eigenvector with zero eigenvalue
by ${\vec \xi}^{(a)}~(a=H,L,h)$,  i.e., ${\bf \Gamma}(p^2=m_a^2){\vec \xi}^{(a)}=0$.
 Then the matrix $U$ that links  $\sigma_i$ and the Higgs $h$
to the mass eigenstates, denoted by $\sigma_H,\sigma_L, h'$,  is given by
\al{
U &=
\pmat{
\xi^{(H)}_1 & \xi^{(L)}_1 &\xi^{(h)}_1\\
\xi^{(H)}_2 & \xi^{(L)}_2 &\xi^{(h)}_2\\
\xi^{(H)}_3 & \xi^{(L)}_3 &\xi^{(h)}_3
},
}
where the canonical dimension of $\xi^{(a)}_1$ and  
$\xi^{(a)}_2$ is one, while that of $\xi^{(a)}_3$ is zero.
This implies 
\al{
\lim_{p^2 \to m^2_a}
{\vec \xi}^{(a)}\Gamma(p^2){\vec \xi}^{(a)}
& = Z_a^{-1}(p^2-m_a^2),
}
and hence
\al{
\pmat{
\sigma_1\\
\sigma_2\\
h
}
&=\pmat{
\xi^{(H)}_1  Z_H^{1/2}& \xi^{(L)}_1  Z_L^{1/2}&\xi^{(h)}_1\\
\xi^{(H)}_2  Z_H^{1/2}& \xi^{(L)}_2 Z_L^{1/2}&\xi^{(h)}_2\\
\xi^{(H)}_3  Z_H^{1/2}& \xi^{(L)}_3 Z_L^{1/2}&\xi^{(h)}_3
}\pmat{
\sigma_H\\
\sigma_L\\
h'
}.
}
The wave function renormalization constants $Z_a$ are dimensionless so that $\sigma_H$, $\sigma_L$ and $h'$ are canonically normalized fields with the canonical  dimension of one.
The Lagrangian \eqref{MFA2ap} is rewritten in terms of the fields $\sigma_H$, $\sigma_L$ and $h'$.

If $m_{\rm DM} (m_{H,L}) >2 M_{1,2}$, $\phi^\pm (\sigma_{H,L})$ would decay into $S_1$ and $S_2$ (the inverse propagators $\Gamma$s have an imaginary part)
within the framework of the effective theory, because the effective theory cannot incorporate confinement.
Therefore, we will consider only the parameter space with $m_{\rm DM}$, $m_{H,L}< 2 M_{1,2}$.

\subsection{Effective interactions}
\begin{figure}
\includegraphics[width=16cm]{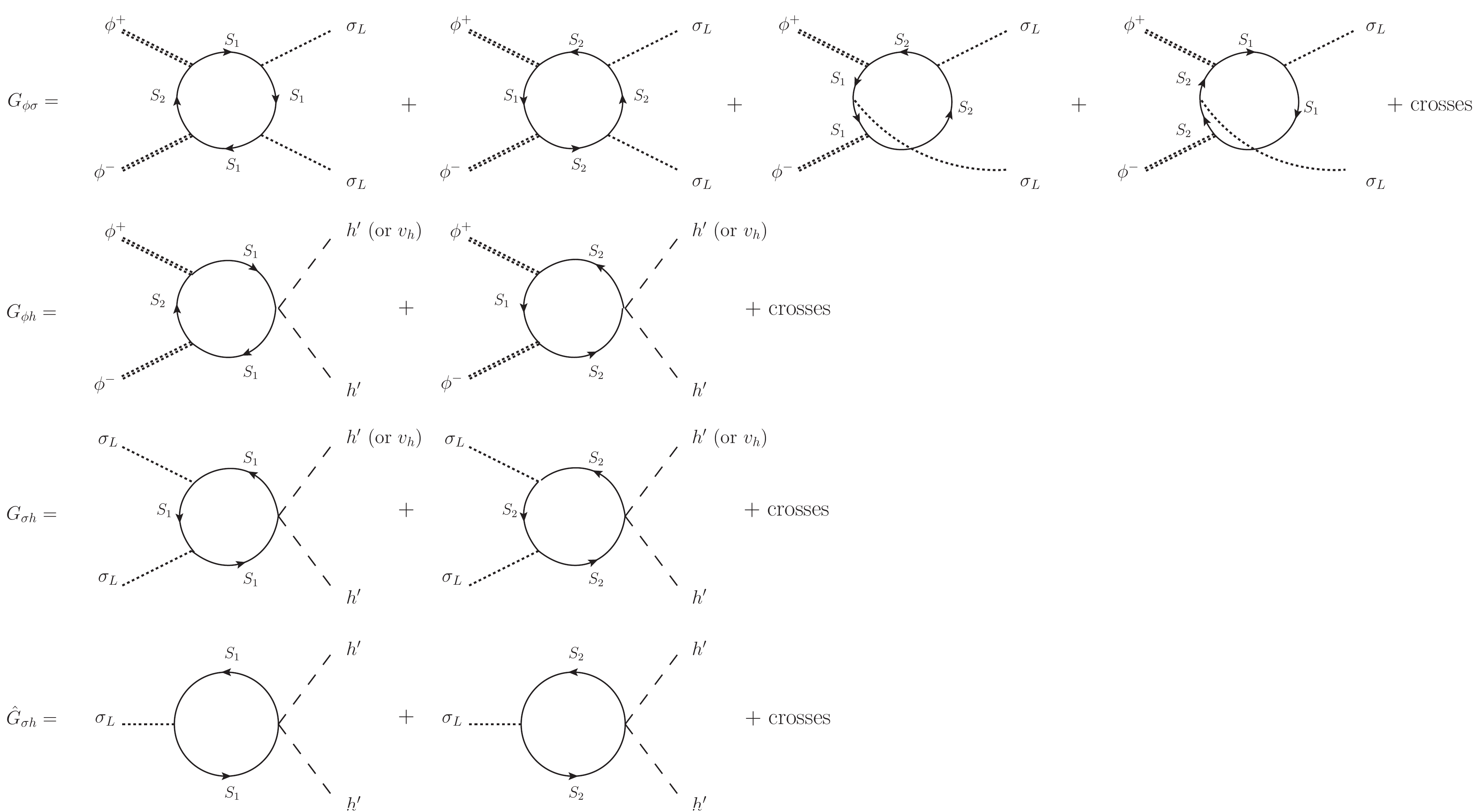}
\caption{\label{effective verticies}
One-loop diagrams contributing to the effective interactions
among $\phi^\pm, \sigma_L$ and $h'$,
where the external momenta are set equal to zero.
}
\end{figure}

To calculate the relic abundance of DM and also
the interaction of DM with the SM particles,
we need to compute the diagrams shown in Fig.\,\ref{effective verticies}.
The corresponding effective interactions can be obtained by setting the external momenta equal to zero:
\al{
{\cal L}_{\rm DM}
&=\frac{1}{2}G_{\phi\sigma}\phi^+\phi^- (\sigma_L)^2+
\frac{1}{2}G_{\phi h}\phi^+\phi^- h'^2+
\frac{1}{4}G_{\sigma h}\sigma_L^2 h'^2\nn
&\quad
+
v_h G_{\phi h}\phi^+\phi^- h'
+\frac{1}{2}v_h G_{\sigma h}\sigma_L^2  h'
+\frac{1}{2}\hat{G}_{\sigma h}\sigma_L  h'^2,
\label{LDM}}
where the effective vertices up to and including 
$O(\lambda_{HS_i})$ are
\al{
G_{\phi\sigma} &= \frac{Z_\phi Z_L N_c}{16 \pi^2} (\lambda'_{12})^2 \left[
\lambda_{1L}^2 F_1(M_1,M_2)+
\lambda_{2L}^2 F_1(M_2,M_1)+
2\lambda_{1L}
\lambda_{2L} F_2(M_1,M_2)\right],
\label{G phisigma coupling}
\\
G_{\phi h} &=   \frac{Z_\phi N_c}{32 \pi^2}  (\lambda'_{12})^2\left[
\lambda_{HS_1} F_3(M_1,M_2)+
\lambda_{HS_2} F_3(M_2,M_1)\right],
\label{G phih coupling}
\\
G_{\sigma h} &=  \frac{Z_L N_c}{16 \pi^2}  \left[
\lambda_{1L}^2\lambda_{HS_1}/M_1^2+
\lambda_{2L}^2\lambda_{HS_2}/M_2^2\right] ,\\
\hat{G}_{\sigma h} &=  \frac{Z_L^{1/2} N_c}{16 \pi^2}  \left[
\lambda_{1L}\lambda_{HS_1}\ln \left(
\frac{M_1^2}{\Lambda_H^2\exp (-3/2)}\right)+
\lambda_{2L}\lambda_{HS_2}
\ln \left(
\frac{M_2^2}{\Lambda_H^2\exp (-3/2)}\right)\right],
\label{G sigmah coupling}
}
with
\al{
\lambda_{1L} &=(2\lambda_1\xi_1^{(L)} +\lambda_{12}\xi_2^{(L)}),&
\lambda_{2L} &=(2\lambda_2\xi_2^{(L)} +\lambda_{12}\xi_1^{(L)}),&
}
and
\al{
F_1(M_1,M_2) &=\frac{M_1^2+M_2^2}{(M_1^2-M_2^2)^2 M_1^2}
-\frac{2M_2^2}{(M_1^2-M_2^2)^3}\ln (M_1^2/M_2^2)&\nn
&= \frac{1}{3M_1^4},~~~\mbox{for}~M_2=M_1,&\\
F_2(M_1,M_2) &=
-\frac{2}{(M_1^2-M_2^2)^2}
+\frac{M_1^2+M_2^2}{(M_1^2-M_2^2)^3}
\ln (M_1^2/M_2^2)&\nn
&= \frac{1}{6M_1^4},~~~\mbox{for}~M_2=M_1,&\\
F_3(M_1,M_2) &=
\frac{1}{M_1^2-M_2^2}
-\frac{M_2^2}{(M_1^2-M_2^2)^2}
\ln (M_1^2/M_2^2)&\nn
&= \frac{1}{2 M_1^2},~~~\mbox{for}~M_2=M_1.&
}
In the next section, the vertices \eqref{G phisigma coupling}--\eqref{G sigmah coupling} are used to evaluate the thermal averaged cross sections for the annihilation processes of $\sigma_L$, $\phi^\pm$ and the decay width of $\sigma_L$.

\section{Dark matter}
\subsection{Relic abundance}
Let us evaluate the relic abundance of the DM candidates in the model.
To this end, we have to follow the temperature-evolution of the number densities of the particles $\sigma_L$ and $\phi^\pm$, denoted by $n_{\sigma_L}$ and $n_\phi$.
These quantities are functions of temperature $T$.
Here, we introduce convenient quantities $Y_{\sigma_L,\phi}=n_{\sigma_L,\phi}/s$, where $s$ is the entropy density of the universe.
Then, the evolution of $Y_{\sigma_L}$ and $Y_\phi$
can be described by the following coupled Boltzmann equation~\cite{DEramo:2010keq,Belanger:2011ww,Belanger:2012vp,Aoki:2012ub}:
\al{
\frac{d Y_{\sigma_L}}{dx}=&
-0.264~ g_*^{1/2} \left[\frac{\mu M_{\rm PL}}{x^2} \right]
\Bigg\{\langle\sigma (\sigma_L\sigma_L;\mbox{SM}) v\rangle
\left(  Y_{\sigma_L}Y _{\sigma_L}-\bar{Y}_{\sigma_L}
\bar{Y}_{\sigma_L}\right)\nn
&\quad
+\langle\sigma (\sigma_L\sigma_L;\phi^+ \phi^- )v\rangle
\left(  Y_{\sigma_L} Y _{\sigma_L}-\frac{Y_{\phi } Y_{\phi }}
{\bar{Y}_{\phi }\bar{Y}_{\phi }} 
\bar{Y}_{\sigma_L}\bar{Y}_{\sigma_L}
\right)\Bigg\}\nn
&\qquad
-0.602~ g_*^{-1/2} \left[\frac{x M_{\rm PL}}{\mu^2} \right]
~ \langle\gamma (\sigma_L)\rangle
  (Y_{\sigma_L}-\bar{Y}_{\sigma_L}),
\label{boltz21}
\\
\frac{d Y_{\phi}}{dx}=&-0.264~ g_*^{1/2} 
\left[\frac{\mu M_{\rm PL}}{x^2} \right]
\Bigg\{ \frac{1}{2}\langle\sigma (\phi^+ \phi^- ;\mbox{SM}) v\rangle
\left(  Y_{\phi} Y _{\phi }-\bar{Y}_{\phi}\bar{Y}_{\phi }\right)\nn
&\quad
-\langle\sigma (\sigma_L\sigma_L;\phi^+ \phi^- )v\rangle
\!\!\left(  Y_{\sigma_L} Y _{\sigma_L}-\frac{Y_{\phi } Y_{\phi}}
{\bar{Y}_{\phi }\bar{Y}_{\phi}} \bar{Y}_{\sigma_L}
\bar{Y}_{\sigma_L}
\right)\Bigg\},
\label{boltz22}
}
where $\bar{Y}_{\sigma_L,\phi}$ is $Y_{\sigma_L,\phi}$ in equilibrium, $M_{\rm PL}=1.22\times  10^{19}~\text{GeV}$ and $g_*=106.75$ are the reduced Planck mass and the total number of effective degrees of freedom, respectively,  and $1/\mu=1/m_L+1/m_{\rm DM}$.
$Y_{\sigma_L,\phi}$ are written as functions of $x=\mu/T$.
Note that $m_{\rm DM}$ is the mass of $\phi^\pm$: $m_\text{DM}\equiv m_\phi$.
The thermal averaged cross sections and the decay width given in \eqref{boltz21} and \eqref{boltz22} are computed as
\al{
 \langle\sigma (\sigma_L\sigma_L;\phi^+ \phi^- ) v\rangle
& =\frac{{G^2_{\phi \sigma}}}{32\pi m_\sigma^2}
\left(1-m_{\rm DM}^2/m_\sigma^2\right)^{1/2},\\
\langle\sigma (\phi^+ \phi^- ;\mbox{SM}) v\rangle
& =\frac{{1}}{32\pi m_{\rm DM}^2}~\sum_{I=W,Z,t,h}
\left(1-m_I^2/m_{\rm DM}^2\right)^{1/2} a_I(G_{\phi h},m_{\rm DM}),\\
 \langle\sigma (\sigma_L\sigma_L;\mbox{SM}) v\rangle
& =\frac{{1}}{32\pi m_L^2}~\sum_{I=W,Z,t,h}
\left(1-m_I^2/m_L^2\right)^{1/2} a_I(G_{\sigma h},m_L),\\
 \langle\gamma (\sigma_L)\rangle &=
 \frac{{1}}{16\pi m_L}~\sum_{I=W,Z,t}
\left(1-4m_I^2/m_L^2\right)^{1/2}  a_I(\hat{G}_{\sigma h},m_L/2)\nn
&\quad + \frac{\hat{G}^2_{\sigma h}}{32\pi m_L}~
\left(1-4m_h^2/m_L^2\right)^{1/2}\left(
1+ 24 \lambda_H \Delta_{h}(m_L/2) \frac{m_W^2}{g^2}\right) ,
}
where $m_W$, $m_Z$, and $m_t$ are the $W$, $Z$ bosons and the top-quark masses given in \eqref{mass at vacuum v}, respectively,  the effective coupling constants are in \eqref{G phisigma coupling}--\eqref{G sigmah coupling}, and we defined
\al{
a_{W(Z)}(G,m)&= 4 (2)
G^2\Delta_{h}^2(m) m_{W(Z)}^4
\left( 3+4\frac{m^4}{m_{W(Z)}^4}-4
\frac{m^2}{m_{W(Z)}^2}\right),\nonumber \\
a_t(G,m)&= 24G^2\Delta_{h}^2 (m) 
m_t^2(m^2-m_t^2), \\
a_h (G,m)&=\frac{1}{2}G^2\left(
1+24 \lambda_H \Delta_{h}(m) \frac{m_W^2}{g^2}+
8G\Delta_{h}^t(m) \frac{m_W^2}{g^2}
\right)^2.\nonumber
}
Here, $g=0.65$ is the $SU(2)_L$ gauge coupling constant,
and $ \Delta_{h}(m)=(4 m^2-m_h^2)^{-1}$ 
($ \Delta_{h}^t(m)=(-2 m^2+m_h^2)^{-1}$)  is
the Higgs propagator in the $s$($t$)-channel.
From the solutions $Y_{\sigma_L;\,\infty}\equiv Y_{\sigma_L}\fn{x=\infty}$ and $Y_{\phi;\,\infty}\equiv Y_\phi\fn{x=\infty}$ to the coupled Boltzmann equation \eqref{boltz21}, \eqref{boltz22}, we obtain the relic abundances for $\sigma_L$ and $\phi^\pm$:
\al{
\Omega_{\sigma_L, \phi}h^2=\frac{g_{\sigma_L,\phi}m_\text{DM} Y_{\sigma_L, \phi;\,\infty} s_0 }{\rho_c/h^2},
}
where $g_{\sigma_L,\phi}$ is the degrees of freedom of $\sigma_L$, $\phi^\pm$, and $s_0=2890\,\text{cm}^{-3}$ and $\rho_c/h^2=1.05\times10^{-5}\,\text{GeV}/\text{cm}^3$ are the entropy density and the critical energy density over the dimensionless Hubble constant at present, respectively~\cite{Patrignani:2016xqp}.

\begin{figure}
\includegraphics[width=8.3cm]{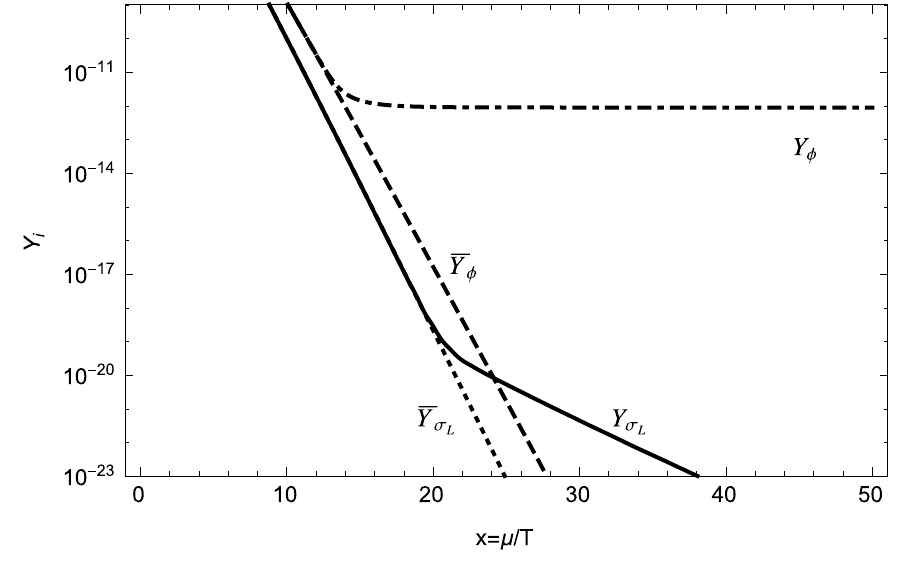}
\includegraphics[width=8cm]{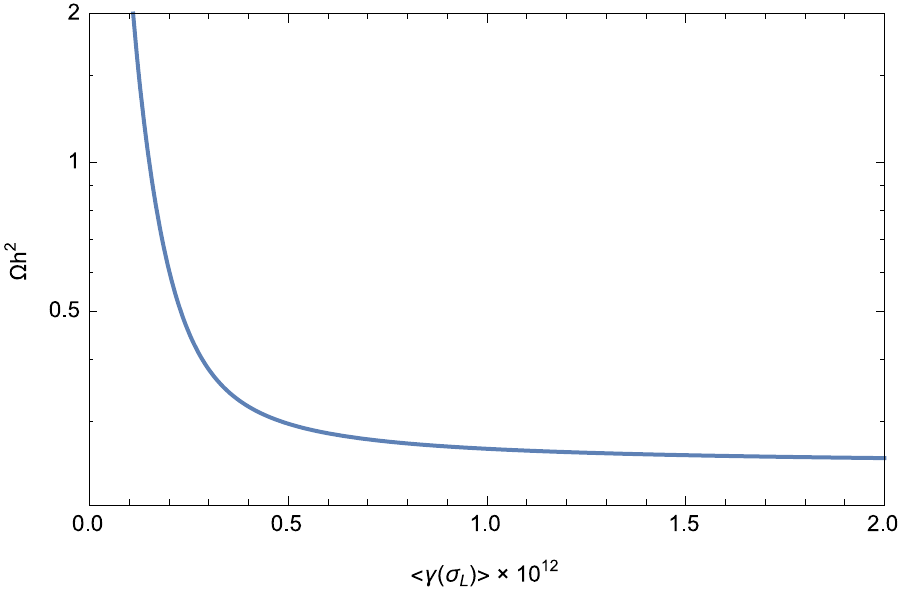}
\caption{\label{yis}
Left: $Y_i$ as a function of $x$.
Right: The total relic abundance $\Omega h^2$
against the decay width $\langle\gamma (\sigma_L)\rangle$
in the range $(0.1\sim 2.0)\times 10^{12}$ GeV.
We have used:
$m_{\rm DM}=500$ GeV, $m_L=550$ GeV,
$\langle\sigma (\sigma_L\sigma_L;\phi^+ \phi^- ) v\rangle
 =5.2\times 10^{-6}\,\mbox{GeV}^{-2}$, $\langle\sigma (\sigma_L\sigma_L;\mbox{SM}) v\rangle=10^{-11}\,\mbox{GeV}^{-2}$, $\langle\sigma (\phi^+ \phi^- ;\mbox{SM}) v\rangle=10^{-11}\,\mbox{GeV}^{-2}$ for both the left- and right-hand panels, while $\langle\gamma (\sigma_L)\rangle=10^{-9}$\,GeV is assumed for the left-hand panel.}
\end{figure}

Before we solve the evolution equations numerically,
we consider what we would expect.
If the decay width $ \langle\gamma (\sigma_L)\rangle$ 
of $\sigma_L$ is large,  $Y_{\sigma_L}$ may be approximated 
by its equilibrium value $\bar{Y}_{\sigma_L}$, which is illustrated
in Fig.\,\ref{yis} for a representative set of the parameters.
From the left-hand side panel of Fig.\,\ref{yis} we see that $Y_{\sigma_L}$ (solid line) can be well approximated by its equilibrium value 
$\bar{Y}_{\sigma_L}$ (dotted line) to compute the final value of for $Y_{\phi}$ (dot-dashed line).
In the right-hand side panel of Fig.\,\ref{yis} we plot the total relic abundance
$\Omega h^2= (\Omega_{\sigma_L}+\Omega_{\phi}) h^2$ 
against the decay width $\langle\gamma (\sigma_L)\rangle$
with the same input parameter (except for $\langle\gamma (\sigma_L)\rangle$) as for the left-hand side panel of Fig.\,\ref{yis},
 where we varied 
 $\langle\gamma (\sigma_L)\rangle$
 between ($0.1$ and $2.0$)$\times 10^{-12}$\,GeV.
 We see that
 the total relic abundance  
approximately  coincides with $\Omega_{\phi} h^2$
if $\langle\gamma (\sigma_L)\rangle \times 10^{12}\,\text{GeV}>0.5$.
Therefore, if  $\langle\gamma (\sigma_L)\rangle$ is sufficiently large,
we may approximate  the expression in the braces $\{~\}$ in the right-hand side of \eqref{boltz22} by
\al{
\left[\frac{1}{2}\langle\sigma (\phi^+ \phi^- ;\mbox{SM}) v\rangle
+\frac{1}{4}\langle\sigma (\sigma_L\sigma_L;\phi^+ \phi^- )v\rangle
\frac{m^3_{\sigma_L}}{m_{\rm DM}^3}
\exp \left(2x\frac{m^2_{\rm DM}
-m^2_L}{m_{\rm DM} m_L}\right) \right]
\left(  Y_{\phi } Y _{\phi }-\bar{Y}_{\phi}\bar{Y}_{\phi}\right),
\label{s-eff}
}
which also appears in the co-annihilation of DM with an unstable 
particle \cite{Griest:1990kh}.
From \eqref{s-eff} we see that if $m_L$ is close to $m_{\rm DM}$ the second term of \eqref{s-eff} 
effectively increases the annihilation rate of DM. 
The reason why $m_L > m_{\rm DM}$ is assumed is that $G_{\phi\sigma}$ given in \eqref{G phisigma coupling} is  so large that the second term in the bracket $[~~]$ of \eqref{s-eff} should be suppressed by $\exp \left(2x\frac{m^2_{\rm DM}
-m^2_L}{m_{\rm DM} m_L}\right)$.
Apart from this mass relation the mechanism is similar to the secluded DM mechanism~\cite{Pospelov:2007mp}.
We use this mechanism\footnote{The decay width 
$\gamma(\sigma_L)$ is typically $\gtrsim O(10^{-10})$ GeV 
in our model.
That is, its lifetime is $\lesssim O(10^{-14} )\,\mbox{s}$, and therefore, the decay of $\sigma_L$ does neither influence BBN nor CMB~\cite{Pospelov:2010hj,Poulin:2016anj}.}
to overcome the constraint from the direct detection experiment, as we explain below.
On one hand, $G_{\phi h}$ enters in the spin-independent 
elastic cross section $\sigma_{SI}$ \eqref{SI}, 
so that it cannot be made small.
The annihilation cross section
 $\langle\sigma (\phi^+ \phi^- ;\mbox{SM}) v\rangle$,
 on the other hand, depends on $G_{\phi h}$, so that there would be a
 lower bound on the relic abundance $\Omega_{\rm DM}$
 of DM,
 if there would be no effect from $\sigma_L$ on $\Omega_{\rm DM}$.
As we have seen above, the $\sigma_L$ effect is an increase
of the annihilation cross section of DM, and consequently, 
the lower bound on $\Omega_{\rm DM}$ can be lowered.

Solving the Boltzmann equation \eqref{boltz22} with the replacement \eqref{s-eff} for large $\langle\gamma (\sigma_L)\rangle$, we obtain the DM relic abundance $\Omega_{\rm DM}h^2$.
The latest observation by the Planck satellite tells us that $\Omega_{\rm DM}h^2= 0.1188\pm 0.0010$~\cite{Ade:2015xua}.

\subsection{Direct detection}
In order to compare with the WIMP DM direct-detection search experiments~\cite{Akerib:2012ys,Aprile:2015uzo,Cao:2014jsa}, we evaluate the spin-independent elastic cross section off the nucleon $\sigma_{SI}$.
As we can see from  ${\cal L}_{\rm DM}$ in (\ref{LDM}) the localized interaction of DM with the SM is that of the Higgs portal.
Consequently, the spin-independent elastic cross section off the nucleon $\sigma_{SI}$ is given by \cite{Barbieri:2006dq}
\al{
\sigma_{SI}
&=\frac{{1}}{4\pi} 
\left(\hat{r} \frac{G_{\phi h} m_N ^2}{m_{\rm DM}m_h^2}
\right)^2
\left(\frac{m_{\rm DM}}{m_N+m_{\rm DM}}
\right)^2,
\label{SI}
}
where $G_{\phi h}$ is given in \eqref{G phih coupling}, $m_N\simeq 940\,\text{MeV}$ is the nucleon mass, and $\hat{r}\sim 0.3$ stems from the nucleonic matrix element~\cite{Ellis:2000ds,Oksuzian:2012rzb,Hoferichter:2015dsa}.
We search the parameter space where the following observed values are satisfied: $v_h=246\,\text{GeV}$, $m_h\simeq 125\,\text{GeV}$, $\Omega_\text{DM}h^2\simeq 0.12$~\cite{Patrignani:2016xqp,Ade:2015xua}.

\begin{figure}
 \includegraphics[width=9cm]{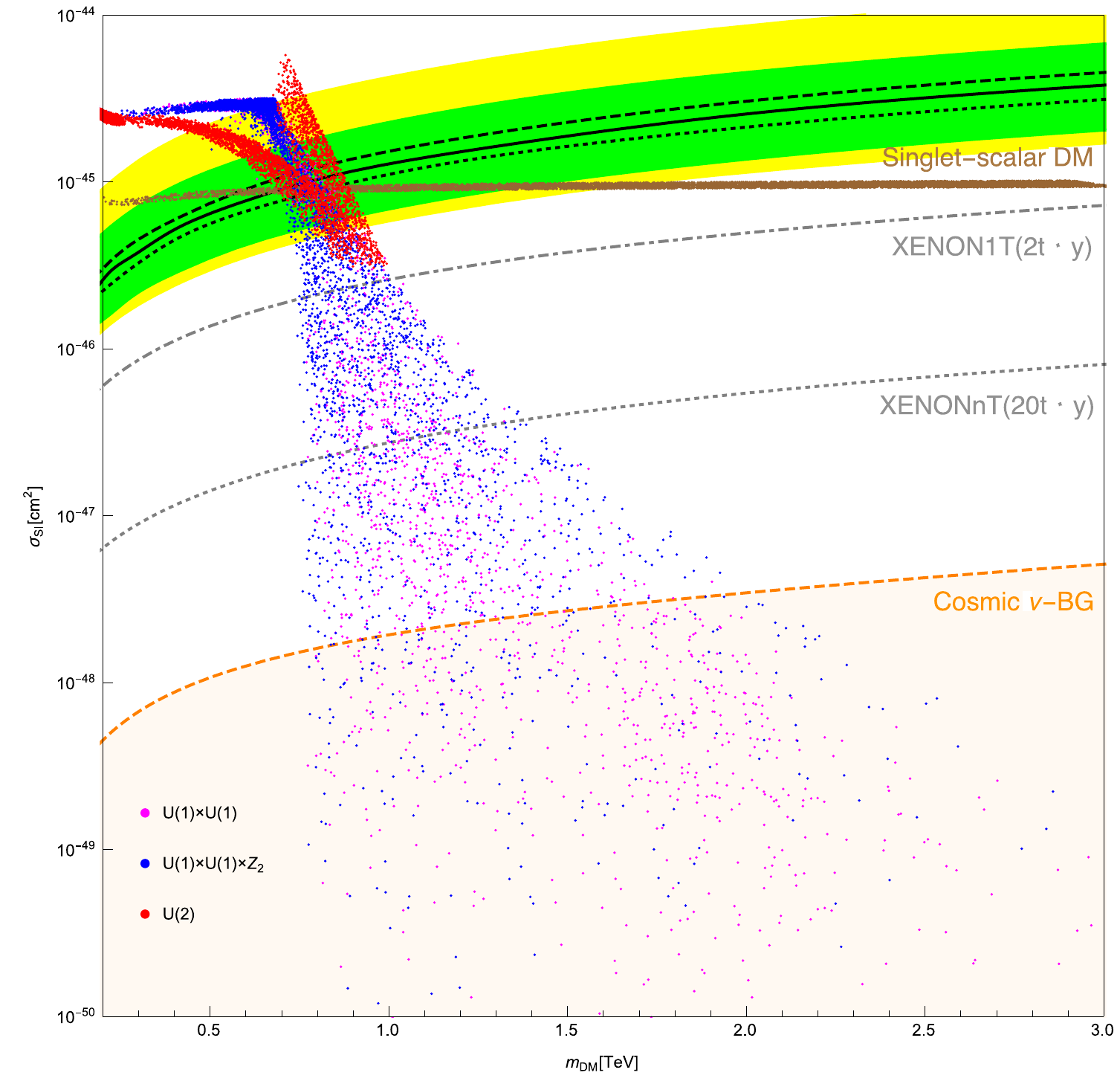}
\caption{\label{sigma-mDM}
The spin-independent elastic cross section $\sigma_\text{SI}$ of DM off the nucleon as a function of the DM mass $m_\text{DM}$ for the case of $N_f=2$, $N_c=6$.
The red, blue and pink points show the predicted regions in
 the model with $U(2)$, $U(1)\times U(1)\times Z_2$ and $U(1)\times U(1)$, respectively.
The brown points show the predicted region
 of the single-scalar DM.
The black dashed, solid and dotted lines stand for the current upper bound from the direct detection experiments, LUX~\cite{Akerib:2016vxi}, XENON1T~\cite{Aprile:2017iyp} and PandaX-II~\cite{Cui:2017nnn}, respectively.
The green and yellow bands denote the $1\sigma$ and $2\sigma$ bands of XENON1T~\cite{Aprile:2017iyp}, respectively.
The gray dot-dashed and dotted lines stands for sensitivities of XENON experiment in the future~\cite{Aprile:2015uzo}.
The orange line and band stands for the cosmic neutrino background~\cite{Billard:2013qya}.
}
\end{figure}

So far we have assumed the $U(1)\times U(1)$ flavor symmetry,
where one of $U(1)$ symmetries is a subgroup of $SU(2)$.
It is possible to enlarge the flavor symmetry, while maintaining the new DM annihilation process, and add the permutation symmetry $Z_2$ of $S_1$ and $S_2$, which requires
 $\lambda_1=\lambda_2$ and $ \lambda_{HS_1}=
 \lambda_{HS_2}$ in ${\cal L}_{\rm eff}$ given in (\ref{Leff}).
 We have computed 
the spin-independent elastic cross section $\sigma_\text{SI}$ of DM off the nucleon for three different flavor symmetries
$U(2), U(1)\times U(1)$ and $U(1)\times U(1)\times Z_2$
with $N_c=6$.
This is shown in Fig.\,\ref{sigma-mDM},
where the red, blue and pink points show the predicted regions in
 the model with $U(2)$, $U(1)\times U(1)\times Z_2$ and $U(1)\times U(1)$, respectively.
For comparison the case of  the single-scalar DM is 
also included (brown points).
These theoretical predictions should be compared with the resent
experimental constraints of LUX~\cite{Akerib:2016vxi}, 
XENON1T~\cite{Aprile:2017iyp} and 
PandaX-II~\cite{Cui:2017nnn}, where
the green and yellow bands denote the $1\sigma$ and $2\sigma$ bands of XENON1T~\cite{Aprile:2017iyp}, respectively.
We see from Fig.\,\ref{sigma-mDM} that the model with 
the unbroken $U(2)$ flavor symmetry is at the border
of the experimental upper bound and future experiments can
exclude the model.
We also see that, in contrast to the $U(2)$ case,
the model  with $U(1)\times U(1)\times Z_2$ and $U(1)\times U(1)$ can clear more stringent constraints.

\section{Conclusion}
We have considered  the scale invariant extension of the SM
proposed in \cite{Kubo:2015cna}, while relaxing the assumption
on the $U(N_f)$ flavor symmetry.
Specifically, we have investigated the model with 
the  $U(2)$ flavor symmetry, which
is broken explicitly down to
$U(1)\times U(1)$ by the scalar quartic couplings.
This breaking opens a completely new possibility
of reducing the relic abundance of DM:
One of the three DM candidates in the $U(2)$ case becomes
neutral under $U(1)\times U(1)$, so that 
the other two ones can annihilate into a pair of the neutral ones,
which subsequently decay in the SM particles.
The result is given in Fig.\,\ref{sigma-mDM}, which shows that the model 
could satisfy  more stringent  constraints  of the future experiments of  DM direct detection. 
A salient feature of the model is that 
 the  DM  of the present model (which is the lightest scalar in the hidden sector)
can be significantly heavier than about $500$ GeV, which is the upper bound for a certain class of classically scale invariant extensions of the SM \cite{Hashino:2015nxa}. 

The  solution of the hierarchy problem
within the framework of the classically scale invariant extension of the SM 
is directly connected to the scale invariance properties of its Planck scale embedding. 
We have assumed the classical 
scale invariance to act in such a way 
that the
Planck scale does not enter as a physical scale into the SM. 
This sounds like a strong assumption, but might be realistic in asymptotically safe gravity which could be one of candidates for quantum gravity~\cite{Oda:2015sma,Wetterich:2016uxm,Hamada:2017rvn,Eichhorn:2017als}.
 
\subsection*{Acknowledgements}
J.\,K. is partially supported by the Grant-in-Aid for Scientific Research (C) from the Japan Society for Promotion of Science (Grant No.16K05315).
Q.\,M.\,B.\,S. is supported by the Directorate General of Resources for Science, Technology and Higher Education Ministry of Research, Technology and Higher Education of Indonesia.
M.\,Y. is supported by the DFG Collaborative Research Centre SFB1225 (ISOQUANT).

\bibliography{refs}

\begin{thebibliography}{57}
\expandafter\ifx\csname natexlab\endcsname\relax\def\natexlab#1{#1}\fi
\expandafter\ifx\csname bibnamefont\endcsname\relax
  \def\bibnamefont#1{#1}\fi
\expandafter\ifx\csname bibfnamefont\endcsname\relax
  \def\bibfnamefont#1{#1}\fi
\expandafter\ifx\csname citenamefont\endcsname\relax
  \def\citenamefont#1{#1}\fi
\expandafter\ifx\csname url\endcsname\relax
  \def\url#1{\texttt{#1}}\fi
\expandafter\ifx\csname urlprefix\endcsname\relax\def\urlprefix{URL }\fi
\providecommand{\bibinfo}[2]{#2}
\providecommand{\eprint}[2][]{\url{#2}}

\bibitem[{\citenamefont{Aad et~al.}(2012)}]{Aad:2012tfa}
\bibinfo{author}{\bibfnamefont{G.}~\bibnamefont{Aad}} \bibnamefont{et~al.}
  (\bibinfo{collaboration}{ATLAS}), \bibinfo{journal}{Phys. Lett.}
  \textbf{\bibinfo{volume}{B716}}, \bibinfo{pages}{1} (\bibinfo{year}{2012}),
  \eprint{1207.7214}.

\bibitem[{\citenamefont{Chatrchyan et~al.}(2012)}]{Chatrchyan:2012xdj}
\bibinfo{author}{\bibfnamefont{S.}~\bibnamefont{Chatrchyan}}
  \bibnamefont{et~al.} (\bibinfo{collaboration}{CMS}), \bibinfo{journal}{Phys.
  Lett.} \textbf{\bibinfo{volume}{B716}}, \bibinfo{pages}{30}
  (\bibinfo{year}{2012}), \eprint{1207.7235}.

\bibitem[{\citenamefont{Bardeen}(1995)}]{Bardeen:1995kv}
\bibinfo{author}{\bibfnamefont{W.~A.} \bibnamefont{Bardeen}}, in
  \emph{\bibinfo{booktitle}{{Ontake Summer Institute on Particle Physics Ontake
  Mountain, Japan, August 27-September 2, 1995}}} (\bibinfo{year}{1995}),
  \urlprefix\url{http://lss.fnal.gov/cgi-bin/find_paper.pl?conf-95-391}.

\bibitem[{\citenamefont{Wetterich}(1984)}]{Wetterich:1983bi}
\bibinfo{author}{\bibfnamefont{C.}~\bibnamefont{Wetterich}},
  \bibinfo{journal}{Phys. Lett.} \textbf{\bibinfo{volume}{140B}},
  \bibinfo{pages}{215} (\bibinfo{year}{1984}).

\bibitem[{\citenamefont{Coleman and Weinberg}(1973)}]{Coleman:1973jx}
\bibinfo{author}{\bibfnamefont{S.~R.} \bibnamefont{Coleman}} \bibnamefont{and}
  \bibinfo{author}{\bibfnamefont{E.~J.} \bibnamefont{Weinberg}},
  \bibinfo{journal}{Phys. Rev.} \textbf{\bibinfo{volume}{D7}},
  \bibinfo{pages}{1888} (\bibinfo{year}{1973}).

\bibitem[{\citenamefont{Hur and Ko}(2011)}]{Hur:2011sv}
\bibinfo{author}{\bibfnamefont{T.}~\bibnamefont{Hur}} \bibnamefont{and}
  \bibinfo{author}{\bibfnamefont{P.}~\bibnamefont{Ko}}, \bibinfo{journal}{Phys.
  Rev. Lett.} \textbf{\bibinfo{volume}{106}}, \bibinfo{pages}{141802}
  (\bibinfo{year}{2011}), \eprint{1103.2571}.

\bibitem[{\citenamefont{Heikinheimo et~al.}(2014)\citenamefont{Heikinheimo,
  Racioppi, Raidal, Spethmann, and Tuominen}}]{Heikinheimo:2013fta}
\bibinfo{author}{\bibfnamefont{M.}~\bibnamefont{Heikinheimo}},
  \bibinfo{author}{\bibfnamefont{A.}~\bibnamefont{Racioppi}},
  \bibinfo{author}{\bibfnamefont{M.}~\bibnamefont{Raidal}},
  \bibinfo{author}{\bibfnamefont{C.}~\bibnamefont{Spethmann}},
  \bibnamefont{and} \bibinfo{author}{\bibfnamefont{K.}~\bibnamefont{Tuominen}},
  \bibinfo{journal}{Mod. Phys. Lett.} \textbf{\bibinfo{volume}{A29}},
  \bibinfo{pages}{1450077} (\bibinfo{year}{2014}), \eprint{1304.7006}.

\bibitem[{\citenamefont{Holthausen et~al.}(2013)\citenamefont{Holthausen, Kubo,
  Lim, and Lindner}}]{Holthausen:2013ota}
\bibinfo{author}{\bibfnamefont{M.}~\bibnamefont{Holthausen}},
  \bibinfo{author}{\bibfnamefont{J.}~\bibnamefont{Kubo}},
  \bibinfo{author}{\bibfnamefont{K.~S.} \bibnamefont{Lim}}, \bibnamefont{and}
  \bibinfo{author}{\bibfnamefont{M.}~\bibnamefont{Lindner}},
  \bibinfo{journal}{JHEP} \textbf{\bibinfo{volume}{12}}, \bibinfo{pages}{076}
  (\bibinfo{year}{2013}), \eprint{1310.4423}.

\bibitem[{\citenamefont{Kubo et~al.}(2014{\natexlab{a}})\citenamefont{Kubo,
  Lim, and Lindner}}]{Kubo:2014ida}
\bibinfo{author}{\bibfnamefont{J.}~\bibnamefont{Kubo}},
  \bibinfo{author}{\bibfnamefont{K.~S.} \bibnamefont{Lim}}, \bibnamefont{and}
  \bibinfo{author}{\bibfnamefont{M.}~\bibnamefont{Lindner}},
  \bibinfo{journal}{JHEP} \textbf{\bibinfo{volume}{09}}, \bibinfo{pages}{016}
  (\bibinfo{year}{2014}{\natexlab{a}}), \eprint{1405.1052}.

\bibitem[{\citenamefont{Heikinheimo and Spethmann}(2014)}]{Heikinheimo:2014xza}
\bibinfo{author}{\bibfnamefont{M.}~\bibnamefont{Heikinheimo}} \bibnamefont{and}
  \bibinfo{author}{\bibfnamefont{C.}~\bibnamefont{Spethmann}},
  \bibinfo{journal}{JHEP} \textbf{\bibinfo{volume}{12}}, \bibinfo{pages}{084}
  (\bibinfo{year}{2014}), \eprint{1410.4842}.

\bibitem[{\citenamefont{Carone and Ramos}(2015)}]{Carone:2015jra}
\bibinfo{author}{\bibfnamefont{C.~D.} \bibnamefont{Carone}} \bibnamefont{and}
  \bibinfo{author}{\bibfnamefont{R.}~\bibnamefont{Ramos}},
  \bibinfo{journal}{Phys. Lett.} \textbf{\bibinfo{volume}{B746}},
  \bibinfo{pages}{424} (\bibinfo{year}{2015}), \eprint{1505.04448}.

\bibitem[{\citenamefont{Ametani et~al.}(2015)\citenamefont{Ametani, Aoki, Goto,
  and Kubo}}]{Ametani:2015jla}
\bibinfo{author}{\bibfnamefont{Y.}~\bibnamefont{Ametani}},
  \bibinfo{author}{\bibfnamefont{M.}~\bibnamefont{Aoki}},
  \bibinfo{author}{\bibfnamefont{H.}~\bibnamefont{Goto}}, \bibnamefont{and}
  \bibinfo{author}{\bibfnamefont{J.}~\bibnamefont{Kubo}},
  \bibinfo{journal}{Phys. Rev.} \textbf{\bibinfo{volume}{D91}},
  \bibinfo{pages}{115007} (\bibinfo{year}{2015}), \eprint{1505.00128}.

\bibitem[{\citenamefont{Haba et~al.}(2016)\citenamefont{Haba, Ishida, Kitazawa,
  and Yamaguchi}}]{Haba:2015qbz}
\bibinfo{author}{\bibfnamefont{N.}~\bibnamefont{Haba}},
  \bibinfo{author}{\bibfnamefont{H.}~\bibnamefont{Ishida}},
  \bibinfo{author}{\bibfnamefont{N.}~\bibnamefont{Kitazawa}}, \bibnamefont{and}
  \bibinfo{author}{\bibfnamefont{Y.}~\bibnamefont{Yamaguchi}},
  \bibinfo{journal}{Phys. Lett.} \textbf{\bibinfo{volume}{B755}},
  \bibinfo{pages}{439} (\bibinfo{year}{2016}), \eprint{1512.05061}.

\bibitem[{\citenamefont{Hatanaka et~al.}(2016)\citenamefont{Hatanaka, Jung, and
  Ko}}]{Hatanaka:2016rek}
\bibinfo{author}{\bibfnamefont{H.}~\bibnamefont{Hatanaka}},
  \bibinfo{author}{\bibfnamefont{D.-W.} \bibnamefont{Jung}}, \bibnamefont{and}
  \bibinfo{author}{\bibfnamefont{P.}~\bibnamefont{Ko}}, \bibinfo{journal}{JHEP}
  \textbf{\bibinfo{volume}{08}}, \bibinfo{pages}{094} (\bibinfo{year}{2016}),
  \eprint{1606.02969}.

\bibitem[{\citenamefont{Ishida et~al.}(2017)\citenamefont{Ishida, Matsuzaki,
  Okawa, and Omura}}]{Ishida:2017ehu}
\bibinfo{author}{\bibfnamefont{H.}~\bibnamefont{Ishida}},
  \bibinfo{author}{\bibfnamefont{S.}~\bibnamefont{Matsuzaki}},
  \bibinfo{author}{\bibfnamefont{S.}~\bibnamefont{Okawa}}, \bibnamefont{and}
  \bibinfo{author}{\bibfnamefont{Y.}~\bibnamefont{Omura}},
  \bibinfo{journal}{Phys. Rev.} \textbf{\bibinfo{volume}{D95}},
  \bibinfo{pages}{075033} (\bibinfo{year}{2017}), \eprint{1701.00598}.

\bibitem[{\citenamefont{Haba and Yamada}(2017{\natexlab{a}})}]{Haba:2017wwn}
\bibinfo{author}{\bibfnamefont{N.}~\bibnamefont{Haba}} \bibnamefont{and}
  \bibinfo{author}{\bibfnamefont{T.}~\bibnamefont{Yamada}},
  \bibinfo{journal}{Phys. Rev.} \textbf{\bibinfo{volume}{D95}},
  \bibinfo{pages}{115016} (\bibinfo{year}{2017}{\natexlab{a}}),
  \eprint{1701.02146}.

\bibitem[{\citenamefont{Haba and Yamada}(2017{\natexlab{b}})}]{Haba:2017quk}
\bibinfo{author}{\bibfnamefont{N.}~\bibnamefont{Haba}} \bibnamefont{and}
  \bibinfo{author}{\bibfnamefont{T.}~\bibnamefont{Yamada}},
  \bibinfo{journal}{Phys. Rev.} \textbf{\bibinfo{volume}{D95}},
  \bibinfo{pages}{115015} (\bibinfo{year}{2017}{\natexlab{b}}),
  \eprint{1703.04235}.

\bibitem[{\citenamefont{Tsumura et~al.}(2017)\citenamefont{Tsumura, Yamada, and
  Yamaguchi}}]{Tsumura:2017knk}
\bibinfo{author}{\bibfnamefont{K.}~\bibnamefont{Tsumura}},
  \bibinfo{author}{\bibfnamefont{M.}~\bibnamefont{Yamada}}, \bibnamefont{and}
  \bibinfo{author}{\bibfnamefont{Y.}~\bibnamefont{Yamaguchi}},
  \bibinfo{journal}{JCAP} \textbf{\bibinfo{volume}{1707}}, \bibinfo{pages}{044}
  (\bibinfo{year}{2017}), \eprint{1704.00219}.

\bibitem[{\citenamefont{Aoki et~al.}(2017)\citenamefont{Aoki, Goto, and
  Kubo}}]{Aoki:2017aws}
\bibinfo{author}{\bibfnamefont{M.}~\bibnamefont{Aoki}},
  \bibinfo{author}{\bibfnamefont{H.}~\bibnamefont{Goto}}, \bibnamefont{and}
  \bibinfo{author}{\bibfnamefont{J.}~\bibnamefont{Kubo}},
  \bibinfo{journal}{Phys. Rev.} \textbf{\bibinfo{volume}{D96}},
  \bibinfo{pages}{075045} (\bibinfo{year}{2017}), \eprint{1709.07572}.

\bibitem[{\citenamefont{Kubo et~al.}(2014{\natexlab{b}})\citenamefont{Kubo,
  Lim, and Lindner}}]{Kubo:2014ova}
\bibinfo{author}{\bibfnamefont{J.}~\bibnamefont{Kubo}},
  \bibinfo{author}{\bibfnamefont{K.~S.} \bibnamefont{Lim}}, \bibnamefont{and}
  \bibinfo{author}{\bibfnamefont{M.}~\bibnamefont{Lindner}},
  \bibinfo{journal}{Phys. Rev. Lett.} \textbf{\bibinfo{volume}{113}},
  \bibinfo{pages}{091604} (\bibinfo{year}{2014}{\natexlab{b}}),
  \eprint{1403.4262}.

\bibitem[{\citenamefont{Kubo and Yamada}(2016{\natexlab{a}})}]{Kubo:2015cna}
\bibinfo{author}{\bibfnamefont{J.}~\bibnamefont{Kubo}} \bibnamefont{and}
  \bibinfo{author}{\bibfnamefont{M.}~\bibnamefont{Yamada}},
  \bibinfo{journal}{Phys. Rev.} \textbf{\bibinfo{volume}{D93}},
  \bibinfo{pages}{075016} (\bibinfo{year}{2016}{\natexlab{a}}),
  \eprint{1505.05971}.

\bibitem[{\citenamefont{Nambu and
  Jona-Lasinio}(1961{\natexlab{a}})}]{Nambu:1961tp}
\bibinfo{author}{\bibfnamefont{Y.}~\bibnamefont{Nambu}} \bibnamefont{and}
  \bibinfo{author}{\bibfnamefont{G.}~\bibnamefont{Jona-Lasinio}},
  \bibinfo{journal}{Phys. Rev.} \textbf{\bibinfo{volume}{122}},
  \bibinfo{pages}{345} (\bibinfo{year}{1961}{\natexlab{a}}).

\bibitem[{\citenamefont{Nambu and
  Jona-Lasinio}(1961{\natexlab{b}})}]{Nambu:1961fr}
\bibinfo{author}{\bibfnamefont{Y.}~\bibnamefont{Nambu}} \bibnamefont{and}
  \bibinfo{author}{\bibfnamefont{G.}~\bibnamefont{Jona-Lasinio}},
  \bibinfo{journal}{Phys. Rev.} \textbf{\bibinfo{volume}{124}},
  \bibinfo{pages}{246} (\bibinfo{year}{1961}{\natexlab{b}}).

\bibitem[{\citenamefont{Aprile et~al.}(2012)}]{Aprile:2012nq}
\bibinfo{author}{\bibfnamefont{E.}~\bibnamefont{Aprile}} \bibnamefont{et~al.}
  (\bibinfo{collaboration}{XENON100}), \bibinfo{journal}{Phys. Rev. Lett.}
  \textbf{\bibinfo{volume}{109}}, \bibinfo{pages}{181301}
  (\bibinfo{year}{2012}), \eprint{1207.5988}.

\bibitem[{\citenamefont{Aprile et~al.}(2013)}]{Aprile:2013doa}
\bibinfo{author}{\bibfnamefont{E.}~\bibnamefont{Aprile}} \bibnamefont{et~al.}
  (\bibinfo{collaboration}{XENON100}), \bibinfo{journal}{Phys. Rev. Lett.}
  \textbf{\bibinfo{volume}{111}}, \bibinfo{pages}{021301}
  (\bibinfo{year}{2013}), \eprint{1301.6620}.

\bibitem[{\citenamefont{Akerib et~al.}(2013)}]{Akerib:2012ys}
\bibinfo{author}{\bibfnamefont{D.~S.} \bibnamefont{Akerib}}
  \bibnamefont{et~al.} (\bibinfo{collaboration}{LUX}), \bibinfo{journal}{Nucl.
  Instrum. Meth.} \textbf{\bibinfo{volume}{A704}}, \bibinfo{pages}{111}
  (\bibinfo{year}{2013}), \eprint{1211.3788}.

\bibitem[{\citenamefont{Aprile et~al.}(2016)}]{Aprile:2015uzo}
\bibinfo{author}{\bibfnamefont{E.}~\bibnamefont{Aprile}} \bibnamefont{et~al.}
  (\bibinfo{collaboration}{XENON}), \bibinfo{journal}{JCAP}
  \textbf{\bibinfo{volume}{1604}}, \bibinfo{pages}{027} (\bibinfo{year}{2016}),
  \eprint{1512.07501}.

\bibitem[{\citenamefont{Cao et~al.}(2014)}]{Cao:2014jsa}
\bibinfo{author}{\bibfnamefont{X.}~\bibnamefont{Cao}} \bibnamefont{et~al.}
  (\bibinfo{collaboration}{PandaX}), \bibinfo{journal}{Sci. China Phys. Mech.
  Astron.} \textbf{\bibinfo{volume}{57}}, \bibinfo{pages}{1476}
  (\bibinfo{year}{2014}), \eprint{1405.2882}.

\bibitem[{\citenamefont{Kubo and Yamada}(2015)}]{Kubo:2015joa}
\bibinfo{author}{\bibfnamefont{J.}~\bibnamefont{Kubo}} \bibnamefont{and}
  \bibinfo{author}{\bibfnamefont{M.}~\bibnamefont{Yamada}},
  \bibinfo{journal}{PTEP} \textbf{\bibinfo{volume}{2015}},
  \bibinfo{pages}{093B01} (\bibinfo{year}{2015}), \eprint{1506.06460}.

\bibitem[{\citenamefont{Kubo and Yamada}(2016{\natexlab{b}})}]{Kubo:2016kpb}
\bibinfo{author}{\bibfnamefont{J.}~\bibnamefont{Kubo}} \bibnamefont{and}
  \bibinfo{author}{\bibfnamefont{M.}~\bibnamefont{Yamada}},
  \bibinfo{journal}{JCAP} \textbf{\bibinfo{volume}{1612}}, \bibinfo{pages}{001}
  (\bibinfo{year}{2016}{\natexlab{b}}), \eprint{1610.02241}.

\bibitem[{\citenamefont{Coleman et~al.}(1974)\citenamefont{Coleman, Jackiw, and
  Politzer}}]{Coleman:1974jh}
\bibinfo{author}{\bibfnamefont{S.~R.} \bibnamefont{Coleman}},
  \bibinfo{author}{\bibfnamefont{R.}~\bibnamefont{Jackiw}}, \bibnamefont{and}
  \bibinfo{author}{\bibfnamefont{H.~D.} \bibnamefont{Politzer}},
  \bibinfo{journal}{Phys. Rev.} \textbf{\bibinfo{volume}{D10}},
  \bibinfo{pages}{2491} (\bibinfo{year}{1974}).

\bibitem[{\citenamefont{Kobayashi and Kugo}(1975)}]{Kobayashi:1975ev}
\bibinfo{author}{\bibfnamefont{M.}~\bibnamefont{Kobayashi}} \bibnamefont{and}
  \bibinfo{author}{\bibfnamefont{T.}~\bibnamefont{Kugo}},
  \bibinfo{journal}{Prog. Theor. Phys.} \textbf{\bibinfo{volume}{54}},
  \bibinfo{pages}{1537} (\bibinfo{year}{1975}).

\bibitem[{\citenamefont{Abbott et~al.}(1976)\citenamefont{Abbott, Kang, and
  Schnitzer}}]{Abbott:1975bn}
\bibinfo{author}{\bibfnamefont{L.~F.} \bibnamefont{Abbott}},
  \bibinfo{author}{\bibfnamefont{J.~S.} \bibnamefont{Kang}}, \bibnamefont{and}
  \bibinfo{author}{\bibfnamefont{H.~J.} \bibnamefont{Schnitzer}},
  \bibinfo{journal}{Phys. Rev.} \textbf{\bibinfo{volume}{D13}},
  \bibinfo{pages}{2212} (\bibinfo{year}{1976}).

\bibitem[{\citenamefont{Bardeen and Moshe}(1983)}]{Bardeen:1983st}
\bibinfo{author}{\bibfnamefont{W.~A.} \bibnamefont{Bardeen}} \bibnamefont{and}
  \bibinfo{author}{\bibfnamefont{M.}~\bibnamefont{Moshe}},
  \bibinfo{journal}{Phys. Rev.} \textbf{\bibinfo{volume}{D28}},
  \bibinfo{pages}{1372} (\bibinfo{year}{1983}).

\bibitem[{\citenamefont{D'Eramo and Thaler}(2010)}]{DEramo:2010keq}
\bibinfo{author}{\bibfnamefont{F.}~\bibnamefont{D'Eramo}} \bibnamefont{and}
  \bibinfo{author}{\bibfnamefont{J.}~\bibnamefont{Thaler}},
  \bibinfo{journal}{JHEP} \textbf{\bibinfo{volume}{06}}, \bibinfo{pages}{109}
  (\bibinfo{year}{2010}), \eprint{1003.5912}.

\bibitem[{\citenamefont{Belanger and Park}(2012)}]{Belanger:2011ww}
\bibinfo{author}{\bibfnamefont{G.}~\bibnamefont{Belanger}} \bibnamefont{and}
  \bibinfo{author}{\bibfnamefont{J.-C.} \bibnamefont{Park}},
  \bibinfo{journal}{JCAP} \textbf{\bibinfo{volume}{1203}}, \bibinfo{pages}{038}
  (\bibinfo{year}{2012}), \eprint{1112.4491}.

\bibitem[{\citenamefont{Belanger et~al.}(2012)\citenamefont{Belanger, Kannike,
  Pukhov, and Raidal}}]{Belanger:2012vp}
\bibinfo{author}{\bibfnamefont{G.}~\bibnamefont{Belanger}},
  \bibinfo{author}{\bibfnamefont{K.}~\bibnamefont{Kannike}},
  \bibinfo{author}{\bibfnamefont{A.}~\bibnamefont{Pukhov}}, \bibnamefont{and}
  \bibinfo{author}{\bibfnamefont{M.}~\bibnamefont{Raidal}},
  \bibinfo{journal}{JCAP} \textbf{\bibinfo{volume}{1204}}, \bibinfo{pages}{010}
  (\bibinfo{year}{2012}), \eprint{1202.2962}.

\bibitem[{\citenamefont{Aoki et~al.}(2012)\citenamefont{Aoki, Duerr, Kubo, and
  Takano}}]{Aoki:2012ub}
\bibinfo{author}{\bibfnamefont{M.}~\bibnamefont{Aoki}},
  \bibinfo{author}{\bibfnamefont{M.}~\bibnamefont{Duerr}},
  \bibinfo{author}{\bibfnamefont{J.}~\bibnamefont{Kubo}}, \bibnamefont{and}
  \bibinfo{author}{\bibfnamefont{H.}~\bibnamefont{Takano}},
  \bibinfo{journal}{Phys. Rev.} \textbf{\bibinfo{volume}{D86}},
  \bibinfo{pages}{076015} (\bibinfo{year}{2012}), \eprint{1207.3318}.

\bibitem[{\citenamefont{Patrignani et~al.}(2016)}]{Patrignani:2016xqp}
\bibinfo{author}{\bibfnamefont{C.}~\bibnamefont{Patrignani}}
  \bibnamefont{et~al.} (\bibinfo{collaboration}{Particle Data Group}),
  \bibinfo{journal}{Chin. Phys.} \textbf{\bibinfo{volume}{C40}},
  \bibinfo{pages}{100001} (\bibinfo{year}{2016}).

\bibitem[{\citenamefont{Griest and Seckel}(1991)}]{Griest:1990kh}
\bibinfo{author}{\bibfnamefont{K.}~\bibnamefont{Griest}} \bibnamefont{and}
  \bibinfo{author}{\bibfnamefont{D.}~\bibnamefont{Seckel}},
  \bibinfo{journal}{Phys. Rev.} \textbf{\bibinfo{volume}{D43}},
  \bibinfo{pages}{3191} (\bibinfo{year}{1991}).

\bibitem[{\citenamefont{Pospelov et~al.}(2008)\citenamefont{Pospelov, Ritz, and
  Voloshin}}]{Pospelov:2007mp}
\bibinfo{author}{\bibfnamefont{M.}~\bibnamefont{Pospelov}},
  \bibinfo{author}{\bibfnamefont{A.}~\bibnamefont{Ritz}}, \bibnamefont{and}
  \bibinfo{author}{\bibfnamefont{M.~B.} \bibnamefont{Voloshin}},
  \bibinfo{journal}{Phys. Lett.} \textbf{\bibinfo{volume}{B662}},
  \bibinfo{pages}{53} (\bibinfo{year}{2008}), \eprint{0711.4866}.

\bibitem[{\citenamefont{Pospelov and Pradler}(2010)}]{Pospelov:2010hj}
\bibinfo{author}{\bibfnamefont{M.}~\bibnamefont{Pospelov}} \bibnamefont{and}
  \bibinfo{author}{\bibfnamefont{J.}~\bibnamefont{Pradler}},
  \bibinfo{journal}{Ann. Rev. Nucl. Part. Sci.} \textbf{\bibinfo{volume}{60}},
  \bibinfo{pages}{539} (\bibinfo{year}{2010}), \eprint{1011.1054}.

\bibitem[{\citenamefont{Poulin et~al.}(2017)\citenamefont{Poulin, Lesgourgues,
  and Serpico}}]{Poulin:2016anj}
\bibinfo{author}{\bibfnamefont{V.}~\bibnamefont{Poulin}},
  \bibinfo{author}{\bibfnamefont{J.}~\bibnamefont{Lesgourgues}},
  \bibnamefont{and} \bibinfo{author}{\bibfnamefont{P.~D.}
  \bibnamefont{Serpico}}, \bibinfo{journal}{JCAP}
  \textbf{\bibinfo{volume}{1703}}, \bibinfo{pages}{043} (\bibinfo{year}{2017}),
  \eprint{1610.10051}.

\bibitem[{\citenamefont{Ade et~al.}(2016)}]{Ade:2015xua}
\bibinfo{author}{\bibfnamefont{P.~A.~R.} \bibnamefont{Ade}}
  \bibnamefont{et~al.} (\bibinfo{collaboration}{Planck}),
  \bibinfo{journal}{Astron. Astrophys.} \textbf{\bibinfo{volume}{594}},
  \bibinfo{pages}{A13} (\bibinfo{year}{2016}), \eprint{1502.01589}.

\bibitem[{\citenamefont{Barbieri et~al.}(2006)\citenamefont{Barbieri, Hall, and
  Rychkov}}]{Barbieri:2006dq}
\bibinfo{author}{\bibfnamefont{R.}~\bibnamefont{Barbieri}},
  \bibinfo{author}{\bibfnamefont{L.~J.} \bibnamefont{Hall}}, \bibnamefont{and}
  \bibinfo{author}{\bibfnamefont{V.~S.} \bibnamefont{Rychkov}},
  \bibinfo{journal}{Phys. Rev.} \textbf{\bibinfo{volume}{D74}},
  \bibinfo{pages}{015007} (\bibinfo{year}{2006}), \eprint{hep-ph/0603188}.

\bibitem[{\citenamefont{Ellis et~al.}(2000)\citenamefont{Ellis, Ferstl, and
  Olive}}]{Ellis:2000ds}
\bibinfo{author}{\bibfnamefont{J.~R.} \bibnamefont{Ellis}},
  \bibinfo{author}{\bibfnamefont{A.}~\bibnamefont{Ferstl}}, \bibnamefont{and}
  \bibinfo{author}{\bibfnamefont{K.~A.} \bibnamefont{Olive}},
  \bibinfo{journal}{Phys. Lett.} \textbf{\bibinfo{volume}{B481}},
  \bibinfo{pages}{304} (\bibinfo{year}{2000}), \eprint{hep-ph/0001005}.

\bibitem[{\citenamefont{Ohki et~al.}(2013)\citenamefont{Ohki, Takeda, Aoki,
  Hashimoto, Kaneko, Matsufuru, Noaki, and Onogi}}]{Oksuzian:2012rzb}
\bibinfo{author}{\bibfnamefont{H.}~\bibnamefont{Ohki}},
  \bibinfo{author}{\bibfnamefont{K.}~\bibnamefont{Takeda}},
  \bibinfo{author}{\bibfnamefont{S.}~\bibnamefont{Aoki}},
  \bibinfo{author}{\bibfnamefont{S.}~\bibnamefont{Hashimoto}},
  \bibinfo{author}{\bibfnamefont{T.}~\bibnamefont{Kaneko}},
  \bibinfo{author}{\bibfnamefont{H.}~\bibnamefont{Matsufuru}},
  \bibinfo{author}{\bibfnamefont{J.}~\bibnamefont{Noaki}}, \bibnamefont{and}
  \bibinfo{author}{\bibfnamefont{T.}~\bibnamefont{Onogi}}
  (\bibinfo{collaboration}{JLQCD}), \bibinfo{journal}{Phys. Rev.}
  \textbf{\bibinfo{volume}{D87}}, \bibinfo{pages}{034509}
  (\bibinfo{year}{2013}), \eprint{1208.4185}.

\bibitem[{\citenamefont{Hoferichter et~al.}(2015)\citenamefont{Hoferichter,
  Ruiz~de Elvira, Kubis, and Mei{\ss}ner}}]{Hoferichter:2015dsa}
\bibinfo{author}{\bibfnamefont{M.}~\bibnamefont{Hoferichter}},
  \bibinfo{author}{\bibfnamefont{J.}~\bibnamefont{Ruiz~de Elvira}},
  \bibinfo{author}{\bibfnamefont{B.}~\bibnamefont{Kubis}}, \bibnamefont{and}
  \bibinfo{author}{\bibfnamefont{U.-G.} \bibnamefont{Mei{\ss}ner}},
  \bibinfo{journal}{Phys. Rev. Lett.} \textbf{\bibinfo{volume}{115}},
  \bibinfo{pages}{092301} (\bibinfo{year}{2015}), \eprint{1506.04142}.

\bibitem[{\citenamefont{Akerib et~al.}(2017)}]{Akerib:2016vxi}
\bibinfo{author}{\bibfnamefont{D.~S.} \bibnamefont{Akerib}}
  \bibnamefont{et~al.} (\bibinfo{collaboration}{LUX}), \bibinfo{journal}{Phys.
  Rev. Lett.} \textbf{\bibinfo{volume}{118}}, \bibinfo{pages}{021303}
  (\bibinfo{year}{2017}), \eprint{1608.07648}.

\bibitem[{\citenamefont{Aprile et~al.}(2017)}]{Aprile:2017iyp}
\bibinfo{author}{\bibfnamefont{E.}~\bibnamefont{Aprile}} \bibnamefont{et~al.}
  (\bibinfo{collaboration}{XENON}), \bibinfo{journal}{Phys. Rev. Lett.}
  \textbf{\bibinfo{volume}{119}}, \bibinfo{pages}{181301}
  (\bibinfo{year}{2017}), \eprint{1705.06655}.

\bibitem[{\citenamefont{Cui et~al.}(2017)}]{Cui:2017nnn}
\bibinfo{author}{\bibfnamefont{X.}~\bibnamefont{Cui}} \bibnamefont{et~al.}
  (\bibinfo{collaboration}{PandaX-II}), \bibinfo{journal}{Phys. Rev. Lett.}
  \textbf{\bibinfo{volume}{119}}, \bibinfo{pages}{181302}
  (\bibinfo{year}{2017}), \eprint{1708.06917}.

\bibitem[{\citenamefont{Billard et~al.}(2014)\citenamefont{Billard, Strigari,
  and Figueroa-Feliciano}}]{Billard:2013qya}
\bibinfo{author}{\bibfnamefont{J.}~\bibnamefont{Billard}},
  \bibinfo{author}{\bibfnamefont{L.}~\bibnamefont{Strigari}}, \bibnamefont{and}
  \bibinfo{author}{\bibfnamefont{E.}~\bibnamefont{Figueroa-Feliciano}},
  \bibinfo{journal}{Phys. Rev.} \textbf{\bibinfo{volume}{D89}},
  \bibinfo{pages}{023524} (\bibinfo{year}{2014}), \eprint{1307.5458}.

\bibitem[{\citenamefont{Hashino et~al.}(2016)\citenamefont{Hashino, Kanemura,
  and Orikasa}}]{Hashino:2015nxa}
\bibinfo{author}{\bibfnamefont{K.}~\bibnamefont{Hashino}},
  \bibinfo{author}{\bibfnamefont{S.}~\bibnamefont{Kanemura}}, \bibnamefont{and}
  \bibinfo{author}{\bibfnamefont{Y.}~\bibnamefont{Orikasa}},
  \bibinfo{journal}{Phys. Lett.} \textbf{\bibinfo{volume}{B752}},
  \bibinfo{pages}{217} (\bibinfo{year}{2016}), \eprint{1508.03245}.

\bibitem[{\citenamefont{Oda and Yamada}(2016)}]{Oda:2015sma}
\bibinfo{author}{\bibfnamefont{K.-y.} \bibnamefont{Oda}} \bibnamefont{and}
  \bibinfo{author}{\bibfnamefont{M.}~\bibnamefont{Yamada}},
  \bibinfo{journal}{Class. Quant. Grav.} \textbf{\bibinfo{volume}{33}},
  \bibinfo{pages}{125011} (\bibinfo{year}{2016}), \eprint{1510.03734}.

\bibitem[{\citenamefont{Wetterich and Yamada}(2017)}]{Wetterich:2016uxm}
\bibinfo{author}{\bibfnamefont{C.}~\bibnamefont{Wetterich}} \bibnamefont{and}
  \bibinfo{author}{\bibfnamefont{M.}~\bibnamefont{Yamada}},
  \bibinfo{journal}{Phys. Lett.} \textbf{\bibinfo{volume}{B770}},
  \bibinfo{pages}{268} (\bibinfo{year}{2017}), \eprint{1612.03069}.

\bibitem[{\citenamefont{Hamada and Yamada}(2017)}]{Hamada:2017rvn}
\bibinfo{author}{\bibfnamefont{Y.}~\bibnamefont{Hamada}} \bibnamefont{and}
  \bibinfo{author}{\bibfnamefont{M.}~\bibnamefont{Yamada}},
  \bibinfo{journal}{JHEP} \textbf{\bibinfo{volume}{08}}, \bibinfo{pages}{070}
  (\bibinfo{year}{2017}), \eprint{1703.09033}.

\bibitem[{\citenamefont{Eichhorn et~al.}(2017)\citenamefont{Eichhorn, Hamada,
  Lumma, and Yamada}}]{Eichhorn:2017als}
\bibinfo{author}{\bibfnamefont{A.}~\bibnamefont{Eichhorn}},
  \bibinfo{author}{\bibfnamefont{Y.}~\bibnamefont{Hamada}},
  \bibinfo{author}{\bibfnamefont{J.}~\bibnamefont{Lumma}}, \bibnamefont{and}
  \bibinfo{author}{\bibfnamefont{M.}~\bibnamefont{Yamada}}
  (\bibinfo{year}{2017}), \eprint{1712.00319}.

\end{thebibliography}
\end{document}